\documentclass[trackchanges]{aastex7}
\usepackage{CJK}
\usepackage{multirow}
\usepackage{booktabs} 
\usepackage{verbatim}

\usepackage{amsmath}

\begin{document}
\begin{CJK*}{UTF8}{gbsn}
\title{Studies on the spin and magnetic inclination evolution of magnetars Swift J1834.9$-$0846 under wind braking}

\author[0000-0003-0325-6426]{B. P. Li (李彪鹏)}
\affiliation{State Key Laboratory of Radio Astronomy and Technology, Xinjiang Astronomical Observatory, Chinese Academy of Sciences, 150, Science-1 Street, Urumqi, Xinjiang, 830011, China}
\affiliation{University of Chinese Academy of Sciences, No. 19 Yuquan Road, Beijing 100049, China}
\email{libiaopeng@xao.ac.cn}

\author[0000-0002-0138-3360]{Z. F. Gao (高志福)}
\affiliation{State Key Laboratory of Radio Astronomy and Technology, Xinjiang Astronomical Observatory, Chinese Academy of Sciences, 150, Science-1 Street, Urumqi, Xinjiang, 830011, China}
\email[show]{zhifugao@xao.ac.cn}

\author[0009-0002-7690-7381]{W. Q. Ma (马文琦)}
\affiliation{State Key Laboratory of Radio Astronomy and Technology, Xinjiang Astronomical Observatory, Chinese Academy of Sciences, 150, Science-1 Street, Urumqi, Xinjiang, 830011, China}
\affiliation{University of Chinese Academy of Sciences, No. 19 Yuquan Road, Beijing 100049, China}
\email{mawenqi@xao.ac.cn}

\author[]{W. F. Zhang (张伟丰)}
\affiliation{State Key Laboratory of Radio Astronomy and Technology, Xinjiang Astronomical Observatory, Chinese Academy of Sciences, 150, Science-1 Street, Urumqi, Xinjiang, 830011, China}
\affiliation{University of Chinese Academy of Sciences, No. 19 Yuquan Road, Beijing 100049, China}
\email{zhangweifeng@xao.ac.cn}

\author[]{Q. Cheng (程泉)}
\affiliation{Institute of Astrophysics, Central China Normal University, Wuhan 430079, China}
\email{qcheng@ccnu.edu.cn}

\author[]{L.C. Garcia de Andrade}
\affiliation{Cosmology and Gravitation Group, Departamento de Física Teórica-IF-UERJ, Rua São Francisco Xavier 524, Maracanã, Rio de Janeiro, RJ, CEP:20550, Brazil}
\affiliation{Institute for Cosmology and Philosophy of Nature, Križevci, Croatia}
\email{luizandra795@gmail.com}

\begin{abstract}

\textbf{The magnetar Swift J1834.9$-$0846 presents a significant challenge to neutron star spin-down models. It exhibits two key anomalies: an insufficient rotational energy loss rate to power its observed X-ray luminosity, and a braking index of $ = 1.08\pm 0.04$,  which starkly  contradicts the canonical magnetic dipole value of $n=3$. To explain these anomalies, we develop a unified spin-evolution model that self-consistently integrates magnetic dipole radiation, gravitational wave emission, and wind braking. Within this framework, we constrain the wind braking parameter to $\kappa \in [13, 37]$ from the nebular properties, finding it contributes substantially ($17\%$--$51\%$) to the current spin-down torque. Bayesian inference reveals that the birth period is poorly constrained by present data and is prior-dependent, indicating a millisecond birth is allowed but not required. Furthermore, we constrain the number of precession cycles to $\xi \sim 10^{4}$--$10^{5}$, and our analysis favors a toroidally-dominated internal magnetic field configuration as the most self-consistent explanation for the low braking index. Finally, we assess the continuous gravitational-wave detectability. The present-day signal is undetectable. However, the early-time signal might have reached the projected sensitivity of next-generation gravitational-wave observatories, such as the Advanced Laser Interferometer Gravitational-Wave Observatory\,(aLIGO) and the Einstein Telescope\,(ET), although a confident detection would require exceptionally stable rotation, an assumption considered highly optimistic for a young magnetar.
This work establishes a unified framework that links magnetar spin-down with their interior physics and multi-messenger observables, providing a physically consistent interpretation for Swift J1834.9$-$0846 and a new tool for understanding similar extreme neutron stars.}

\end{abstract}


\keywords{
\uat{Magnetars}{992} --- 
\uat{Neutron stars}{1108} --- 
\uat{Pulsar wind nebulae}{2215} --- 
\uat{Magnetic fields}{994} --- 
\uat{Gravitational waves}{678}
}


\section{Introduction}
\label{sec:intro}
Magnetars are a class of young, isolated neutron stars  characterized by ultra-strong magnetic fields, typically exceeding $10^{14}$\,G. Based on their high-energy emission properties, they are usually classified into two categories: soft gamma-ray repeaters (SGRs) and anomalous X-ray pulsars (AXPs). These objects display a rich variety of X-ray phenomena, including bursts, flares, and intriguing timing irregularities such as glitches and anti-glitches\,\citep{Kaspi2017ARAA}. The predominant energy source for their high-energy activity is thought to be the decay of their internal magnetic fields. To date,  observations have confirmed or identified roughly thirty magnetar candidates\,\citep{Olausen2014ApJS}. Current research endeavors are primarily focused on understanding the origin and evolution of their magnetic fields, the physical mechanisms driving their high-energy outbursts, and their long-term rotational evolution, including potential links to fast radio bursts (FRBs)\,\citep{Mereghetti2015SSRv, Thompson2004ApJ, Turolla2015RPPh, Bochenek2020Natur}. 

The magnetar Swift J1834.9$-$0846 (hereafter Swift J1834) serves as a crucial test case for understanding 
these phenomena. It was discovered during an outburst in 2011\,\citep{DElia2011GCN, 
Goldstein2012ApJS}, subsequent observations with RXTE and Chandra have determined the spin period of $P = 2.48\,\mathrm{s}$ and the period derivative $\dot{P} = 0.796\times10^{-11}\,\mathrm{s\,s^{-1}}$ \citep{Kuiper2011ATel, Gogus2011ATel}. These values imply a surface dipole field $B_{\mathrm{surf}}=1.4\times10^{14}$\,G, a spin-down luminosity $\dot{E}_{\mathrm{rot}}=2.1\times10^{34}\,\mathrm{erg\,s^{-1}}$ and a characteristic age  $t_{\mathrm{c}}=4.9\times10^{3}\,\mathrm{yrs}$. Swift J1834 is positionally coincident with the geometric center of the supernova remnant\,(SNR) W41\,\citep{Kargaltsev2012ApJ}. \cite{Torres2017ApJ} discussed the possible association between Swift J1834 and W41. While definitive proof remains elusive, the precise alignment at the SNR center ($\Delta\theta < 0.1'$,\,\citealt{Tian2007ApJ, Leahy2008AJ}) strongly suggests a potential physical connection. The estimated age of SNR W41 is $t_{\mathrm{SNR}} = 130 \pm 70$ kyr. This potential association, and the significant age discrepancy it presents if real 
($\tau_c \ll t_{\mathrm{SNR}}$), provides an initial, intriguing puzzle for its evolutionary history. However, as no independent, conclusive age constraint exists, our subsequent analysis in Section\,\ref{sec:Inference of Initial Parameters} will treat the magnetar's age as a free parameter to be constrained by its current timing properties within our spin-down model, allowing the data to self-consistently determine the most probable evolutionary path.

The braking index ($n \equiv \Omega\ddot{\Omega}/\dot{\Omega}^2$, where $\Omega$ is the angular velocity, $\dot{\Omega}$ its time derivative representing the spin-down rate, and $\ddot{\Omega}$ the second time derivative quantifying the temporal evolution of spin-down)  remains a crucial parameter for understanding neutron star rotational evolution. The standard magnetic dipole radiation\,(MDR) model predicts $n=3$. However, observations systematically reveal deviations from this canonical value. Although over 3000 pulsars are known, only eight have a well-measured, stable braking index, and none are consistent with the pure dipole prediction\,\citep{Espinoza2011MNRAS, Lyne1993MNRAS, Livingstone2007ApSS,Weltevred2011MNRAS, Lyne1996Natur,Roy2012MNRAS,Ferdman2015ApJ}. These discrepancies have motivated the development of alternative spin-down mechanisms, among which gravitational wave emission (GWE), capable of producing $n > 3$, and wind braking, which typically yields $n < 3$, are particularly important candidates\,\citep{Ostriker1969ApJ,Araujo2016JCAP,Kou2015MNRAS,Xu2001ApJ}. Swift J1834 presents an striking case study for these phenomena. The comprehensive analysis by\,\citet{Gao2016MNRAS} combining its precise timing data with SNR W41's age estimate led to the revelation of an exceptionally low braking index of $n=1.08\pm0.04$. This extreme deviation, significantly lower than typical pulsar values, strongly suggests wind braking dominates the spin-down process in this source. 

The wind braking hypothesis finds direct support in Swift J1834's nebular properties. Similar to ordinary pulsars, magnetars are expected to drive particle outflows-either continuously or impulsively during bursts and flares. This gives rise to extended nebula-like emission structures around magnetars. Two deep XMM-Newton observations carried out after the 2011 outburst confirmed the presence of such a nebula around Swift J1834 \cite{Younes2016ApJ}. This nebula exhibits remarkable energetics: it radiates $1.5\times10^{34}\,\mathrm{erg}\,\mathrm{s}^{-1}$ in 0.5-8 keV, corresponding to an efficiency of roughly seventy per cent with respect to the spin-down luminosity\,\citep{Younes2012ApJ}. Such a high conversion efficiency offers compelling, independent evidence that wind braking must be a major component in any realistic spin-evolution model for this magnetar.

To achieve a comprehensive understanding of magnetar spin-down mechanisms, it is essential to consider three intrinsically coupled physical processes: magnetic field evolution, inclination angle dynamics, and structural deformation. Rotational evolution of neutron stars depends critically on their internal physics and external torques. Spin-down models incorporating the evolution of key internal parameters (including the magnetic field, the magnetic inclination angle, and the moment of inertia) naturally account for observed deviations from the canonical braking index\,\citep{Spitkovsky2006ApJ,Eksi2016ApJ,Gao2017ApJ}. The magnetic fields of neutron stars decay through Ohmic dissipation and Hall drift, while ultra-strong fields induce non-axisymmetric quadrupolar deformations, leading to continuous gravitational wave emission\,\citep{Bonazzola1996AA,Cutler2002PhRvD,Riles2023LRR}.

The complex interplay between electromagnetic and gravitational torques governs the system's orientation evolution. Both magnetic dipole and gravitational wave radiation torques act to align the magnetic and spin axes over time\,\citep{Cutler2000PhRvD, Philippov2014MNRAS}. In highly magnetized stars, the deformation symmetry axis typically misaligns with the spin axis, inducing free precession with period $P_{\mathrm{prec}}=P/\epsilon_{\mathrm{B}}$, where $\epsilon_{\mathrm{B}}$ is the ellipticity due to magnetic deformation. During this process, viscous dissipation inside the star damps the precession, driving the inclination angle toward the minimum-energy configuration determined by the internal magnetic field geometry\,\citep{Alpar1988ApJ, Cutler2002PhRvD}. The damping rate is characterized by the number of precession cycles $\xi \equiv \tau_{\mathrm{DIS}}/P_{\mathrm{prec}}$ \citet{Cutler2002PhRvD}, where $\tau_{\mathrm{DIS}}$ represents the viscous dissipation timescale. Current theoretical estimates for $\xi$ span a wide range ($10 - 10^8$) due to uncertainties in neutron star interior microphysics\,\citep{Alpar1988ApJ,Jones1976ApSS,Haskell2008MNRAS,Cheng2019PhRvD, Hu2023RAA}. This parameter characterizes the specific viscous mechanism and plays a crucial role in discussions of continuous gravitational wave emission, as it determines the timescale for the star to evolve toward the optimal (unfavorable) configuration for gravitational wave emission.

Our work builds upon this theoretical foundation while introducing crucial innovations. We develop a comprehensive spin evolution framework that self-consistently incorporates wind braking effects with magnetic inclination angle evolution, enabling us to place joint constraints on both the precession cycle parameter $\xi$ and the internal magnetic field geometry 
of Swift~J1834.
The paper's organization reflects this systematic methodology. Section \ref{sec: model} develops the complete theoretical framework that integrates wind braking with existing spin-down mechanisms. Section \ref{sec:Results} presents our numerical results, including detailed parameter constraints derived from observational data. Section \ref{sec: Gravitational Wave} explores the gravitational wave detectability predictions based on our model. Finally, Section \ref{sec: Conclusion} summarizes our key findings and discusses their broader implications for understanding magnetar physics and neutron star evolution.

\section{Theoretical Model}
\label{sec: model}
Building on the observational constraints for Swift J1834 summarized in Section~\ref{sec:intro}, notably its anomalously low braking index and prominent wind nebula, we construct a self-consistent spin-down framework. To quantitatively describe its long-term evolution, we develop a model that simultaneously and coherently incorporates the torque contributions from MDR, GWE and wind braking. The central role of wind braking, powered by the magnetar's ultra-strong magnetic field efficiently accelerating particles, is key to understanding the anomalous braking index.

Crucially, this model accounts for the coupled evolution of the key parameters linking these mechanisms, primarily the 
magnetic inclination angle $\chi$ and the dipole magnetic field strength $B_\mathrm{d}$, which are governed by their own differential equations. This approach allows us to move beyond simple torque prescriptions and model the long-term 
evolutionary path of the magnetar from its birth to the present day.

\subsection{Rotational Evolution}
\label{sec:Rotational Evolution}

The rotational evolution of magnetars is governed by the combined effects of multiple torque mechanisms. For Swift J1834, we develop a comprehensive model that simultaneously accounts for magnetic dipole radiation, gravitational wave emission, and wind braking. The total spin-down torque can be expressed as:
\begin{equation}
\dot{\Omega} = \dot{\Omega}_{\text{MDR}} + \dot{\Omega}_{\text{GWE}} + \dot{\Omega}_{\text{wind}}.
\label{eq:total_torque}
\end{equation}
Here, $\dot{\Omega}_{\text{MDR}}$ represents the conventional spin-down torque due to magnetic dipole radiation, which 
provides the dominant braking mechanism for most pulsars. The term $\dot{\Omega}_{\text{GWE}}$ accounts for the energy loss 
through gravitational wave emission, which becomes significant in the presence of a substantial asymmetry (e.g., a large ellipticity $\epsilon_\mathrm{B}$). The term $\dot{\Omega}_{\text{wind}}$ represents the 
additional torque arising from the particle wind, a mechanism that is particularly relevant for magnetars like Swift J1834 due to their ultra-strong magnetic fields and observed wind nebula; this component is primarily responsible for producing low braking indices $n < 3$.

\subsubsection{Wind Braking Model}
\label{sec:Wind Braking}
The wind braking model represents a substantial revision to standard neutron star spin-down theory, introducing a dual-channel energy-loss framework. In magnetars, the mechanism proceeds along two separate but coupled routes: (i) 
coherent magnetic-dipole radiation produced by the perpendicular component of the star's external field, and (ii) incoherent particle acceleration along open field lines. 

\begin{figure}[h]
\centering
\includegraphics[width=0.5\linewidth]{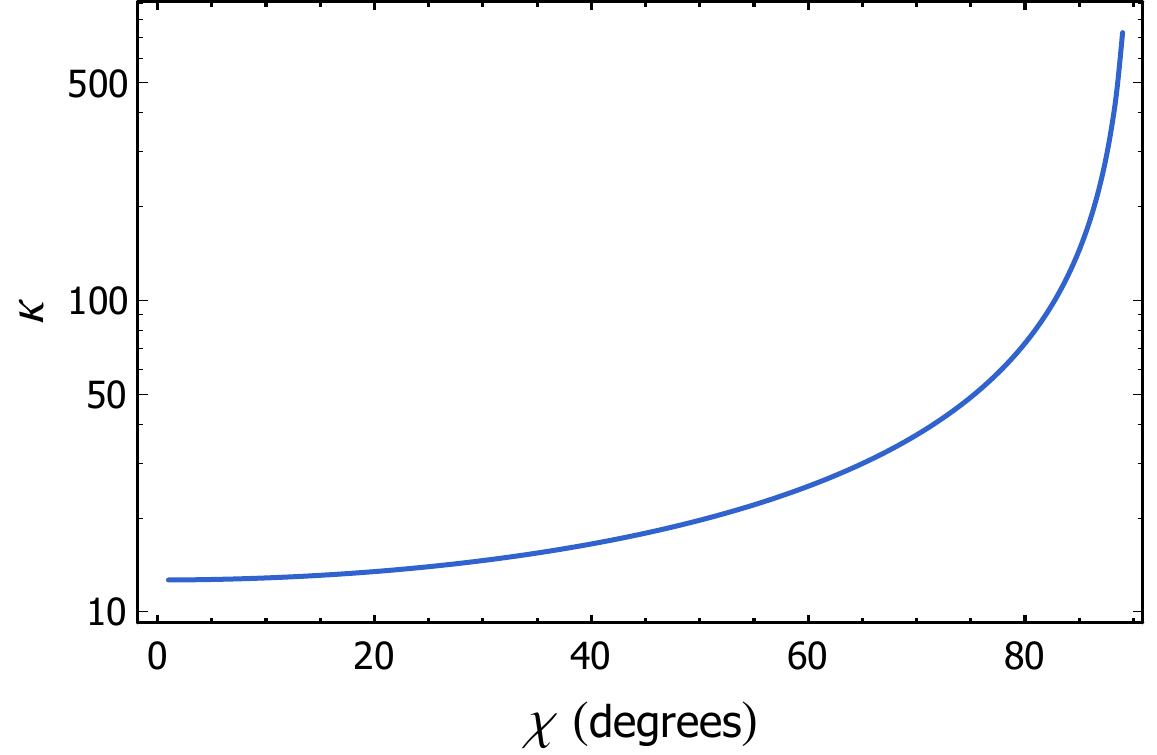}
\caption{The $\kappa$--$\chi$ relation for Swift J1834, obtained from Equation\,(\ref{eq:dew}) and (\ref{eq_eta}) under the assumption that the wind luminosity equals the rotational energy-loss rate.}
\label{fig:kappa_chi}
\end{figure}

Primary particles (electron-positron pairs) are generated through quantum electrodynamic processes in the polar cap, where the energy per particle is determined by the accelerating potential drop $\Delta \nu$\,\citep{Kou2015MNRAS,Tong2017ApJ}.
The rotational energy loss rate derived by \,\citep{Xu2001ApJ}:
\begin{equation}
\dot{E}_{\mathrm{wind}} = 2\pi r_p^2 c \rho_e \Delta \nu,
\label{eq1}
\end{equation}
where $r_p = R(R\Omega/c)^{1/2}$ is the polar gap radius, $\rho_e$ is the primary particle density. The maximum potential drop of a rotating dipole is $ \Delta V = \mu \Omega^2 / c^2$, leading to the normalized spin-down power:
\begin{equation}
\dot{E}_{\mathrm{wind}} = \frac{2\mu^2 \Omega^4}{3c^3} 3\kappa \frac{\Delta \nu}{\Delta V} \cos^2 \chi
= \frac{2\mu^2 \Omega^4}{3c^3}\eta_{\mathrm{wind}},
\label{eq:dew}
\end{equation}
where $\mu=1/2B_{\mathrm{d}}R^3$ is the magnetic dipole moment and $\eta_{\mathrm{wind}}$ is a dimensionless function representing the effective torque and its explicit form depends on the adopted particle acceleration model. Here $\kappa=n_e/n_{\mathrm{GJ}}$ is a dimensionless parameter related to the primary particle density and $n_{\mathrm{GJ}}=\Omega B/2\pi c e$ is the Goldreich-Julian charge density. This factor characterizes the pair production efficiency in the acceleration region, typically ranging from $10^2$--$10^4$ for normal pulsars\,\citep{Ou2016MNRAS, Yue2007AdSpR, Kou2015MNRAS}. 
It is worth noting that the inferred $\kappa$ for magnetars, as we will show for Swift J1834, can be systematically lower due to their distinct magnetospheric physics and pair-production processes in ultra-strong magnetic fields.
According to theoretical models\,\citep{Xu2001ApJ}, only the vacuum gap\,(VG) model with curvature radiation\,(CR), the space charge-limited flow\,(SCLF) model without field saturation, and the outer gap\,(OG) models can yield braking indices as low as observed for Swift J1834. For this source, CR in the vacuum gap model provides the most physically consistent description\,\citep{Ruderman1975ApJ, Kou2015MNRAS}:

\begin{equation}
    \eta_{\mathrm{VG}}^{\mathrm{CR}}=\begin{cases}& 4.96\times10^2\kappa(1-\frac{\Omega_{\mathrm{death}}}{\Omega})B_{12}^{-8/7}\Omega^{-15/7}\cos^{2}{\chi},~~~\mathrm{if~}\Omega>\Omega_{\mathrm{death}}\\&0,~~~\mathrm{if~}\Omega<\Omega_{\mathrm{death}}\end{cases}
    \label{eq_eta}
\end{equation}
where $\Omega_{\mathrm{death}}$ marks the cessation of pair production. The wind braking parameter $\eta_{\mathrm{wind}}$ depends on both $\kappa$ and $\chi$ through Equation\,(\ref{eq_eta}). 
Figure\,\ref{fig:kappa_chi} shows the derived relationship between $\kappa$ and $\chi$ under the assumption that the wind luminosity equals the rotational energy loss rate.

In Figure\,\ref{fig:kappa_chi}, the monotonically increasing trend reflects how a larger $\chi$ implies a smaller $\cos^{2}\chi$ factor; hence, a higher $\kappa$ is needed to keep the wind torque constant, so that the observed spin-down rate is maintained. In realistic magnetar environments, numerical studies of pair cascades show that the $\kappa$ generally decreases as the dipole field $B(t)$ decays\,\citep{Kou2015MNRAS, Cerutti2015MNRAS}. The falling magnetospheric voltage reduces the pair-creation rate, so the particle outflow and wind torque weaken. This gradual decline in $\kappa$ contributes to the spread of observed braking indices by enhancing wind braking in the early, high-field phase and diminishing it at late times, thereby providing a natural link between magnetar wind activity and magnetic-field evolution.

The wind parameter $\kappa$ will not remain constant throughout the magnetar’s lifetime. As the magnetic field decays, the pair-production rate and magnetospheric current density are expected to decrease, leading to a gradual reduction of the wind luminosity. Incorporating this dependence would improve the predictive power of the model in explaining long-term spin evolution.

The determination of magnetic inclination angles in magnetars remains observationally challenging due to limited geometric constraints. So far, only a few magnetars have had their emission geometry constrained via radio polarization or X-ray profile modeling (see Table\,\ref{tab:1}). A notable case is XTE 1810$-$197, where \citet{Camilo2007ApJ} analyzed the pulse profile morphology and polarization data to 
distinguish between two possible geometric configurations: (i) a significantly misaligned magnetic 
and spin axis with high-altitude emission, and (ii) a nearly aligned geometry where the line of sight remains within the 
emission region for almost the entire phase. The inferred inclination angles for these two scenarios are $70^{\circ}$ and 
$4 ^{\circ}$, respectively. The large modulation depth of the thermal X-ray flux definitively 
excluded the nearly aligned scenario, establishing $\chi=70^{\circ}$ as the correct solution.

\begin{table}[h]
\caption{Observed magnetic inclination angle $\chi$ measurements for magnetars.}
\centering
\begin{tabular}{lll}
\hline
Magnetar & $\chi$ (deg) & Reference \\
\hline
1E 1547.0$-$5408 & $<40$ & \citet{Camilo2008ApJ}\\
Swift J1818.0$-$1607 & $68^{+9}_{-7}$ & \citet{Lower2021MNRAS} \\
PSR J1622$-$4950 & $20-46$ & \citet{Levin2012MNRAS} \\
XTE 1810$-$197 & $70$ & \citet{Camilo2007ApJ} \\
1E 2259+586 & $48^{+6}_{-9}$ & \citet{Peng2024ApJ} \\
& $43.1^{+7.0}_{-8.5}$ & \citet{Heyl2024MNRAS}\\
& $21.2^{+4.6}_{-5.0}$ (Mode-switching) & \citet{Heyl2024MNRAS} \\
1RXS J170849.0$-$400910 & $10$ & \citet{Zane2023ApJ} \\
\hline
\end{tabular}
\label{tab:1}
\end{table}

For Swift J1834, we establish the inclination angle range through systematic comparison with these calibrated sources. Constraining the physical parameters of Swift J1834 requires careful incorporation of observational limits drawn from analogous magnetars\,\citep{Olausen2014ApJS}. Because no direct geometric measurement is available for this source, we establish the allowed inclination angle interval through a comprehensive survey of the published literature. Adopting 
$10^{\circ} < \chi < 70^{\circ}$ systematically spans the full range of well-determined magnetar geometries, from the 
nearly aligned rotator 1RXS J170849.0$-$400910 ($\chi = 10^{\circ}$)\,\citep{Zane2023ApJ} to the highly inclined XTE J1810$-$197 ($\chi = 70^{\circ}$)\,\citep{Camilo2007ApJ}. 
By applying this $\chi$ range to the $\kappa-\chi$ relation\,(Figure\,\ref{fig:kappa_chi}), we obtain $\kappa = 
13-37$. This narrow parameter window reflects several aspects of magnetar physics that clearly distinguish these objects from rotation-powered pulsars\,(RPPs)\,\citep{Harding2017ApJ}. 

The inferred $\kappa$ interval for Swift J1834 significantly differs from typical RPP values. Three distinct physical factors contribute to this difference:

1. Ultra-strong surface fields ($B \sim 10^{14}$ G)  substantially alter pair-production thresholds compared to RPPs\,\citep{Thompson1995MNRAS,Vigano2013MNRAS}. 
 
2. The unique magnetospheric topology of magnetars modifies accelerator-gap physics, particularly in the polar-cap regions\,\citep{Turolla2015RPPh,Harding2002ApJ,Medin2007MNRAS}.  
   
3. Relativistic beaming effects are particularly pronounced in magnetar magnetospheres, further altering the effective particle-outflow properties\,\citep{Philippov2022ARAA,Spitkovsky2006ApJ,Cerutti2015MNRAS}.  

Together these effects explain why $\kappa$ in magnetars is systematically lower than in RPPs\,\citep{Harding2017ApJ,Kalapotharakos2018ApJ}.
The present analysis builds on crucial observational studies of magnetar geometry, particularly pulse-profile modeling\,\citep{Younes2022ApJ} and polarimetry\,\citep{Zane2023ApJ}, which provide the empirical basis for our inclination angle bounds. Contemporary magnetospheric theory then guides the interpretation of the wind parameters\,\citep{Philippov2022ARAA,Mereghetti2015SSRv}. Recent population syntheses further confirm the systematic contrast between magnetar and pulsar wind characteristics, especially regarding particle-acceleration efficiency\,\citep{Philippov2022ARAA,Kalapotharakos2018ApJ}.
\subsubsection{Correction to the Braking Index}

Building on \cite{Cheng2019PhRvD}'s framework, we extend the torque model to include wind effect\,\citep{Philippov2014MNRAS,Cheng2019PhRvD,Cutler2000PhRvD,Xu2001ApJ}:

\begin{equation}
\frac{d\Omega}{dt} = -\frac{B_d^2 R^6\Omega^3}{4Ic^3}(1+1.4\sin^2\chi) - \frac{2G\epsilon_B^2 I\Omega^5}{5c^5}
\sin^2\chi(1+15\sin^2\chi) - \frac{B_d^2 R^6\Omega^3}{6Ic^3}\eta_{\mathrm{VG}}^{\mathrm{CR}},
\label{eq:domega}
\end{equation}
where $G$ is the gravitational constant. The three terms on the right-hand side correspond to magnetic dipole radiation, gravitational wave emission, and wind braking, respectively. This equation forms the basis for the numerical simulations presented in Section\,\ref{sec:Results}. We adopt $I=10^{45}\,\mathrm{g}\,\mathrm{cm}^2$ for the stellar moment of inertia and $R=10\,\mathrm{km}$ for the neutron star radius. The second derivative of $\Omega$ is

\begin{equation}
\begin{aligned}
    \frac{d^2\Omega}{dt^2}&=-\bigg(\frac{2\dot{B_{\mathrm{d}}}}{B_{\mathrm{d}}}+\frac{3\dot{\Omega}}{\Omega}+\frac{1.4\dot{\chi}\sin{2\chi}}{1+1.4\sin^{2}{\chi}}\bigg)\frac{B_d^2 R^6\Omega^3}{4Ic^3}(1+1.4\sin^2\chi)\\
    &\quad-\bigg[\frac{5\dot{\Omega}}{\Omega}+\frac{2\dot{\chi}(1+30\sin^{2}{\chi})}{\tan \chi(1+15\sin^2\chi)}\bigg]\frac{2G\epsilon_B^2 I\Omega^5}{5c^5}\sin^2\chi(1+15\sin^2\chi)\\
    &\quad-\bigg(\frac{6\dot{B_{\mathrm{d}}}}{7B_{\mathrm{d}}}+\frac{6\dot{\Omega}}{7\dot{\Omega}}-{2\dot{\chi}\tan{\chi}}\bigg)\frac{B_d^2 R^6\Omega^3}{6Ic^3}\eta_{\mathrm{VG}}^{\mathrm{CR}}.
\end{aligned}  
\label{eq:ddomega}
\end{equation}

By substituting Equations~(\ref{eq:domega}) and~(\ref{eq:ddomega}) into the expression for the braking index, we obtain
\begin{equation}
    \begin{aligned}
    n &= \frac{\gamma}{\eta(1+\gamma)+\gamma}\bigg\{3\eta+\frac{6\eta}{7\gamma}+5+\frac{2\Omega}{\dot{\Omega}}\bigg[\frac{\dot{B}}{B}(\eta+\frac{3\eta}{7\gamma})\\
    & \quad+\dot{\chi}\sin\chi\cos\chi\bigg(\frac{\eta}{1+\sin^2\chi}+\frac{1+30\sin^2\chi}{\sin^2\chi\left(1+15\sin^2\chi\right)}
     \bigg)\bigg]\bigg\},
     \label{eq:n}
    \end{aligned}
\end{equation}
where $\eta={\dot{\Omega}_{\mathrm{MDR}}}/{\dot{\Omega}_{\mathrm{GWE}}}={5B^2_{\mathrm{d}}c^2 R^6(1+1.4\sin^2{\chi})}/[{8I^2 G \epsilon^2_{\mathrm{B}} \Omega^2 \sin^2{\chi}(1+15\sin^2{\chi})}]$ and $\gamma={\dot{\Omega}_{\mathrm{MDR}}}/{\dot{\Omega}_{\mathrm{wind}}}={3(1+1.4\sin^2{\chi})}/{2\eta^{\mathrm{CR}}_{\mathrm{VG}}} $ denote, respectively, the ratios of the MDR spin-down rate to the GWE and wind braking spin-down rates. For Swift J1834, when $10^{\circ}<\chi<70^{\circ}$ is adopted, we obtain $\eta\sim 10^{11}-10^{14}$ and $\gamma\sim 0.1-1.1$. This analysis demonstrates that gravitational wave braking contributes less than 0.1\% of the total spin-down torque at the present epoch, while wind braking becomes significant at $30-50\%$ efficiency, with magnetic dipole radiation dominating the remaining $50-70\%$ contribution. 

\textbf{At the present epoch, the GWE torque is negligible in the spin evolution, but it can still influence the long-term evolution of $\chi$ in our model. By contrast, in the early, rapidly rotating phase, the GWE term can become much more important for the spin evolution. The gravitational waves from the birth phase of magnetars may be detectable.} Wind braking contributes at a level comparable to dipole radiation, and both are crucial for the rotation evolution of magnetars.

\textbf{The model offers a natural explanation for the unusually low braking index observed in Swift J1834. The present-day spin-down is dominated by a combination of magnetic dipole radiation and particle wind, with only a tiny contribution from gravitational waves. In this regime, the effective braking index is already reduced below $3$ by the additional wind component, and is further lowered by the field-decay term and the inclination-angle correction. For a broad region of parameter space compatible with the observational constraints, these combined effects reproduce the measured value $n = 1.08 \pm 0.04$ without requiring fine tuning.}

\textbf{This interpretation is consistent with current theoretical and observational understanding of magnetar magnetospheres. Global magnetospheric simulations indicate that current sheets and particle-loaded outflows can enhance the spin-down torque beyond the vacuum dipole value \citep{Spitkovsky2006ApJ}, while outburst-driven magnetospheric untwisting \citep{Younes2022ApJ} and magnetic-field decay on Hall-drift timescales \citep{Vigano2013MNRAS} provide natural mechanisms for time-dependent $B_{\mathrm{d}}$ and particle loading. In our phenomenological framework, these processes are encoded in the effective wind and field-decay terms that enter Equation~(\ref{eq:n}), and together they offer a coherent explanation for the anomalously low braking index of Swift J1834.}

\subsection{Evolution of the Magnetic Inclination Angle}

The magnetic inclination angle $\chi$ is a key parameter governing the spin-down torque and gravitational wave emission of neutron stars. The evolution of $\chi$ is driven by the combined effects of electromagnetic torques, gravitational wave emission, and internal viscous dissipation. As established in Section\,\ref{sec:intro}, both MDR and GWE contribute to the spin-down of Swift J1834 while simultaneously driving a secular decrease in its magnetic inclination angle $\chi$. Superimposed on this evolution, free precession of the 
neutron star undergoes damping by internal viscous dissipation, leading to additional $\chi$-modulation. Integrating these effects, \citet{Cheng2019PhRvD} derived the magnetic inclination angle evolution equation：

\begin{equation}
\begin{aligned}
    \frac{d\chi}{dt}=\left.\begin{cases}
        &-\frac{B_{d}^2 R^6 \Omega^{2}}{4I c^{3}}\sin{\chi}\cos{\chi}-\frac{2G \epsilon_{B}^{2} I \Omega^{4}}{5c^{5}}\sin \chi\cos\chi(1+15\sin^2\chi)-\frac{\epsilon_B}{\xi P}\tan\chi,~~ \text{for}~~ \epsilon_B>0\\
        &-\frac{B_{d}^2 R^6 \Omega^{2}}{4I c^{3}}\sin{\chi}\cos{\chi}-\frac{2G \epsilon_{B}^{2} I \Omega^{4}}{5c^{5}}\sin \chi\cos\chi(1+15\sin^2\chi)-\frac{\epsilon_B}{\xi P}\cot\chi,~~ \text{for}~~ \epsilon_B<0
    \end{cases}\right.
\end{aligned} 
\label{eq:dchi}
\end{equation}
The first and second terms on the right-hand side describe $\chi$-evolution from MDR and GWE, respectively, while the third term represents viscous damping. The direction of viscous damping (alignment or counter-alignment)  depends critically on the sign of the magnetically induced ellipticity $\epsilon_{\mathrm{B}}$. 

The ellipticity $\epsilon_\mathrm{B}$ depends on the neutron star equation of state, and internal magnetic field configuration. It determines whether the star is oblate ($\epsilon_\mathrm{B} > 0$) or prolate 
($\epsilon_\mathrm{B} < 0$), which in turn governs the evolution of $\chi$ towards alignment or orthogonalization. Following \citet{Cheng2019PhRvD} and \citet{Hu2023RAA}, we adopt $\epsilon_{\mathrm{B}}$ expressions 
incorporating proton superconductivity effects:  

1. \textbf{Poloidal-Dominated (PD) Field: } 

   For superconducting stars with mixed poloidal-toroidal fields\,\citep{Lander2013PhRvL}, the magnetic deformation is  
\begin{equation}
    \epsilon_\mathrm{B} = 3.4 \times 10^{-7} \left( \frac{B_\mathrm{d}}{10^{13}\,\mathrm{G}} \right) \left( \frac{H_{\mathrm{c1}}(0)}{10^{16}\,\mathrm{G}} \right),
    \label{eq:ep_pd}
\end{equation}
where the central critical field strength is taken as $H_{\mathrm{c1}}(0) = 10^{16}$\,G. For Swift J1834, this yields $\epsilon_{\mathrm{B}} \sim 10^{-6}$, corresponding to an oblate deformation. From Equation\,(\ref{eq:dchi}), viscous damping synergizes with MDR and GWE to drive $\chi$ toward spin-magnetic axis alignment ($\chi \rightarrow 0^\circ$). 

2. \textbf{Toroidal-Dominated (TD) Field}:  

   For purely toroidal configurations \citep{Akgun2008MNRAS, Glampedakis2010MNRAS}, we use  
\begin{equation}
    \epsilon_\mathrm{B} \approx -10^{-8} \left( \frac{H}{10^{15}\,\mathrm{G}} \right) \left( \frac{\bar{B}_\mathrm{in}}{10^{13}\,\mathrm{G}} \right),
    \label{eq:ep_td}
\end{equation}
where the critical field strength $H \sim 10^{15}$\,G, and $\bar{B}_\mathrm{in}$ denotes the volume-averaged intensity of the internal toroidal field, which we assume to satisfy $\bar{B}_\mathrm{in} \approx 10B_{\mathrm{d}}$\,\citep{Pons2007PhRvL}. For Swift J1834 with $B_\mathrm{d} \approx 1.4 \times 10^{14} \, \text{G}$, this gives $\epsilon_{\mathrm{B}} \sim -10^{-6}$, producing a prolate deformation. Here, viscous dissipation counteracts the $\chi$-reduction tendency from MDR and GWE, driving the system toward an orthogonal 
configuration ($\chi \rightarrow 90^\circ$).

As detailed in Section\,\ref{sec:omega_chi_Evolution}, the TD configuration in Swift J1834 is favored by the observational data, particularly the low braking index and wind nebula properties. This configuration drives $\chi$ toward orthogonality, thereby maximizing continuous gravitational wave emission. Crucially, the viscous damping parameter $\xi$ in Equation\,(\ref{eq:dchi}) governs the timescale to reach this optimal state, which we constrain observationally in Section\,\ref{sec:constration_xi} through.

\subsection{Evolution of the Magnetic Field}

The evolution of neutron star magnetic fields is a complex process involving multiple physical mechanisms that operate on different timescales. For Swift J1834, we consider the relative importance of these mechanisms in its evolutionary stage.

The temporal evolution of neutron star magnetic fields arises from three primary physical mechanisms: Ohmic dissipation, Hall drift, and ambipolar diffusion. Each mechanism dominates under different conditions and timescales:

\begin{enumerate}
    \item \textbf{Ohmic dissipation} arises from the finite electrical conductivity in the neutron star 
    crust, characterized by the timescale\,\citep{Goldreich1992ApJ,Cumming2004ApJ}:
    \begin{equation}
    \tau_{\text{Ohm}} = \frac{4\pi\sigma_{\|} L^2}{c^2},
    \label{eq:ohmic_timescale}
    \end{equation}
    where $\sigma_{\|}$ is the electrical conductivity parallel to the magnetic field, and $L$ is the characteristic length 
  scale of field variation, typically the crustal thickness.
    
    \item \textbf{Hall drift}, while conserving magnetic energy, enhances the Ohmic dissipation rate by generating 
 small-scale field structures. The Hall timescale is given by\,\citep{Goldreich1992ApJ,Cumming2004ApJ}:
    \begin{equation}
    \tau_{\text{Hall}} = \frac{4\pi n_e e L^2}{c B},
    \label{eq:hall_timescale}
    \end{equation}
    where $n_e$ is the electron number density.
    
    \item \textbf{Ambipolar diffusion} describes the co-motion of magnetic flux and charged particles 
   through the neutron-dominated medium. The governing equation expresses the force balance as \citet{Goldreich1992ApJ}:
    \begin{equation}
    \frac{f_B}{n_c} - \nabla(\Delta\mu) = \left(\frac{m_p}{\tau_{pn}} + \frac{m_e^*}{\tau_{en}}\right) v,
    \label{eq:ambipolar_equation}
    \end{equation}
  where $f_B = j\times B/c$ represents the Lorentz force density, $\Delta\mu = \mu_p + \mu_e - \mu_n$ is the chemical potential gradient, and $v$ denotes the drift velocity.
\end{enumerate}

\textbf{Ambipolar diffusion in the core is expected to play only a minor role for Swift J1834 at its age. Classical estimates for normal $npe$ matter show that the ambipolar drift timescale scales roughly as $\tau_{\mathrm{amb}} \propto L^{2} T^{2} / B^{2}$ and becomes very long once the core cools below a few $\times 10^{8}$\,K \citep{Goldreich1992ApJ}. When neutron superfluidity and proton superconductivity are taken into account, mutual friction together with the suppression of particle collisions and reactions further increase $\tau_{\mathrm{amb}}$, and it is unlikely that standard ambipolar diffusion can drive appreciable magnetic-field evolution in magnetar cores on $\sim 10^{3}$--$10^{4}$\,yrs time-scales \citep{Glampedakis2011MNRAS,Passamonti2017MNRAS,Kantor2018MNRAS}.
}

\textbf{We therefore simplify the magnetic-field evolution model by including only Ohmic dissipation and Hall drift in the crust, which are expected to dominate the observable evolution of the external dipole in strongly magnetized neutron stars \citep{Jones1988MNRAS,Vigano2013MNRAS}.}
The effective decay timescale combines both processes through\,\citep{Gao2017ApJ}:
\begin{equation}
\tau_{\text{eff}}^{-1} = \tau_{\text{Hall}}^{-1} + \tau_{\text{Ohm}}^{-1}.
\label{eq:effective_timescale}
\end{equation}

\textbf{To describe the long-term evolution of the dipole field in a way that incorporates the early, Hall-modulated phase and the later, Ohmic-dominated relaxation, we adopt a analytic prescription\,\citep{Aguilera2008AA, Popov2012MNRAS, Zhou2024MNRAS}:}

\begin{equation}
B_{\mathrm{d}}(t)=\frac{B_0 \exp(-t/\tau_{\mathrm{Ohm}})}{1+(\tau_{\mathrm{Ohm}}/\tau_{\mathrm{Hall}})(1-\exp(-t/\tau_{\mathrm{Ohm}}))},
\label{eq:field_decay}
\end{equation}
\textbf{where $B_0$ is the initial dipole magnetic field strength, and $\tau_{\mathrm{Ohm}}$ and $\tau_{\mathrm{Hall}}$ are the Ohmic and Hall timescales defined above. In realistic neutron-star crusts, Ohmic dissipation and Hall drift are strongly coupled: the Hall term redistributes magnetic energy among different spatial scales, enhancing the local current density and thereby accelerating Ohmic decay. Numerical simulations show that this coupling produces a non-exponential decay characterized by an initially faster Hall-modulated phase and a slower, Ohmic-like relaxation at later times\,\citep{Pons2007PhRvL, Vigano2013MNRAS}. }

\textbf{The analytic model of Equation\,(\ref{eq:field_decay}) provides a smoothed description that encapsulates this two-phase evolution. This model is particularly appropriate for modeling the long-term evolution of middle-aged magnetars like Swift J1834, which has likely passed the initial rapid Hall-dominated phase and resides in the regime where the combined effect of both processes sets the effective decay rate.}

In our numerical implementation, we adopt characteristic values based on neutron star physics:
\begin{itemize}
    \item Observations of magnetic field-temperature correlations suggest $\tau_{\text{Ohm}} \sim 5 
\times 10^5 - 10^6$\,yrs\,\citep{Pons2007PhRvL}
    \item Simulations incorporating core-crust coupling yield $\tau_{\text{Ohm}} \sim 1.5 \times 10^8$ \,yrs\,\citep{Bransgrove2018MNRAS}
    \item For Hall drift, taking the local pressure scale height as the characteristic length, typical 
timescales are $\tau_\mathrm{Hall}=1.2\times 10^4(10^{15} \mathrm{G}/ B_\mathrm{d})$\,yrs\,\citep{Cumming2004ApJ}
\end{itemize}

For Swift J1834 with $B_\mathrm{d} \simeq 1.4\times10^{14}$\,G, this estimate gives 
$\tau_{\text{Hall}}\approx 8.6\times10^4$\,yrs, comparable to or shorter than typical Ohmic timescales. This suggests that Hall-driven evolution likely plays a significant role in its magnetic field history.
This treatment provides a computationally tractable framework while capturing the essential physics of magnetic field evolution in middle-aged magnetars like Swift J1834.

\textbf{It is important to note that the model described by Equation\,(\ref{eq:field_decay}) provides a monotonic approximation. In reality, the strong coupling between Hall drift and Ohmic dissipation can lead to more complex, 
non-monotonic evolutionary paths \citep{Vigano2013MNRAS}. In young, highly magnetized neutron stars where the Hall timescale is short, magnetic energy can be redistributed in ways that cause temporary enhancements or oscillations of the external dipole field $B_\mathrm{d}$, deviating from a purely decaying trend. Furthermore, eruptive events like magnetar outbursts can lead to sudden magnetospheric reconfigurations that may also transiently affect the observed dipole moment \citep{Younes2016ApJ}. While 
our adopted model offers a robust baseline for studying secular evolution, the potential for such non-monotonic behavior introduces an important uncertainty when interpreting the spin-down history and inferring the internal magnetic field configuration. The impact of these complexities on our specific results, particularly regarding the preference for a toroidal-dominated field, will be discussed in Section~\ref{sec: Conclusion}.}

\textbf{The coupled differential equations for the angular velocity $\Omega(t)$ (governed by the total torque in Equation\,(\ref{eq:domega})), the magnetic inclination angle $\chi(t)$ (Equation\,(\ref{eq:dchi})), and the dipole magnetic field $B_\mathrm{d}(t)$ (\ref{eq:field_decay})) form a self-consistent system that defines our theoretical model. This framework describes the rotational and evolutionary history of a magnetar by linking the instantaneous spin-down torque to the long-term evolution of the underlying physical parameters. In the following section, we will numerically solve this coupled system to reconstruct the evolutionary history of Swift J1834 and constrain its initial parameters.}

\setcounter{footnote}{0}
\section{Results}
\label{sec:Results}
\textbf{This section presents the results of our numerical simulations and Bayesian inference, which collectively chart 
the evolutionary path of Swift J1834 from its birth to the present day. We proceed in three steps: first, we show 
case the forward evolution of the magnetar's spin and magnetic inclination based on our theoretical framework 
(Section\,\ref{sec:omega_chi_Evolution});  second, we employ Bayesian analysis to invert this process, constraining the probable initial conditions that 
led to the current observed state (Section\,\ref{sec:Inference of Initial Parameters}); finally, we combine these constraints to probe the magnetar's internal physics, specifically the number of precession cycles $\xi$ and the geometry of the internal magnetic field (Section\,\ref{sec:constration_xi}). 
Together, these results provide a self-consistent reconstruction of the magnetar's history and its interior properties.}

\subsection{Rotational and Magnetic Inclination Evolution of Swift J1834}
\label{sec:omega_chi_Evolution}
This section presents the numerical simulation results of the coupled evolution of the angular velocity 
$\Omega$ and magnetic inclination angle $\chi$ for Swift J1834, based on the theoretical framework developed in Section\,\ref{sec: model}. 
The evolution is governed by the coupled system of Equation\,(\ref{eq:domega}) (spin-down torque) and Equation\,(\ref{eq:dchi}) (inclination angle dynamics), which self-consistently incorporates magnetic dipole radiation, gravitational wave emission, and wind braking 
effects. Our simulations aim to reconstruct the long-term rotational history of Swift J1834 from its birth to the present day, providing insights into the dominant torque mechanisms and their interplay with the internal magnetic field geometry.

We numerically solve the coupled differential equations, with initial conditions set to explore a range of plausible birth scenarios: initial 
spin period $P_0 = 0.02~\text{s}$ and 
initial magnetic inclination angles $\chi_0 = 10^\circ$ and $45^\circ$. The model parameters are chosen as follows: precession cycle parameter $\xi = 10^5$ and wind parameter $\kappa = 25$. We adopt a neutron star moment of inertia $I = 10^{45}~\text{g}~\text{cm}^{2}$ and radius $R = 10~\text{km}$ which are standard values for canonical neutron stars.

\begin{figure}[h]
\centering
\includegraphics[width=0.7\linewidth]{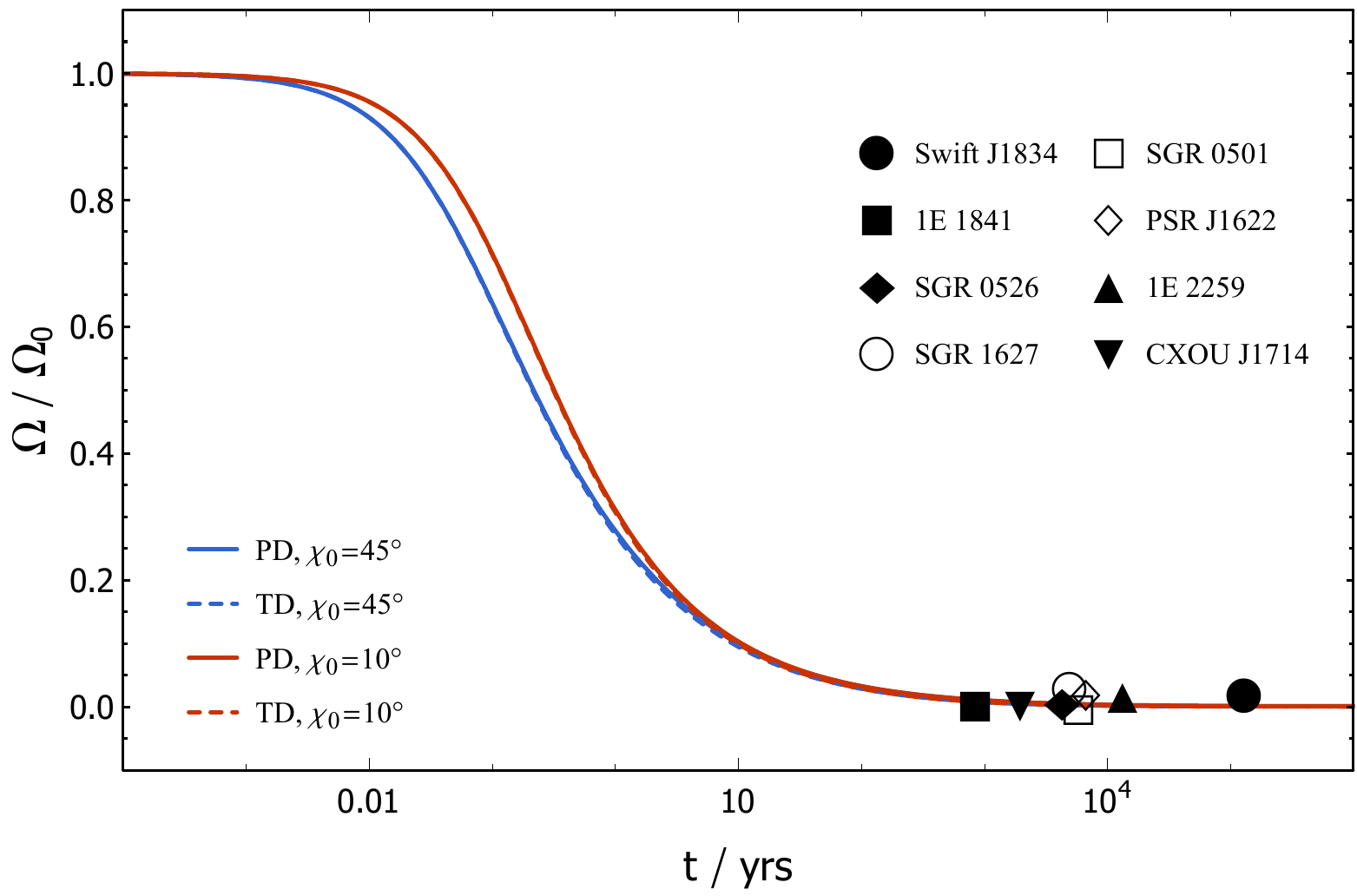}
\caption{Evolution of angular velocity $\Omega$ with time for Swift J1834. Dashed lines: toroidal-dominated (TD) 
internal magnetic configuration. Solid lines: poloidal-dominated (PD) configuration. Red and blue lines correspond to initial inclination angles $\chi_0 = 10^\circ$ and $45^\circ$, respectively. The black dot marks the current observed position of Swift J1834. Back symbols represent other magnetars with supernova remnant age constraints (data from 
\citealt{Gao2016MNRAS}).}
\label{fig:omega_t}
\end{figure}

Figure\,\ref{fig:omega_t} shows the time evolution of the angular velocity $\Omega(t)$ for Swift J1834 under two internal magnetic field configurations: toroidal-dominated (TD, dashed lines) and poloidal-dominated (PD, solid lines). The spin-down behavior exhibits strong nonlinearity, characterized by an initial rapid deceleration phase followed by a gradual approach 
to the present-day value. This nonlinearity is primarily driven by wind braking, which contributes significantly to the total torque at early times when the rotational velocity is high. The evolutionary curves in the TD and PD cases are similar, but different initial magnetic inclinations lead to a clear divergence in the early stages of evolution. \textbf{The divergence of the evolutionary tracks for different initial inclinations $\chi_0$ in Figure\,\ref{fig:omega_t} during the early phase highlights the sensitivity of the spin-down history to the birth geometry, which is subsequently damped by the efficient braking.}

\begin{figure}[h]
\centering
\includegraphics[width=0.7\linewidth]{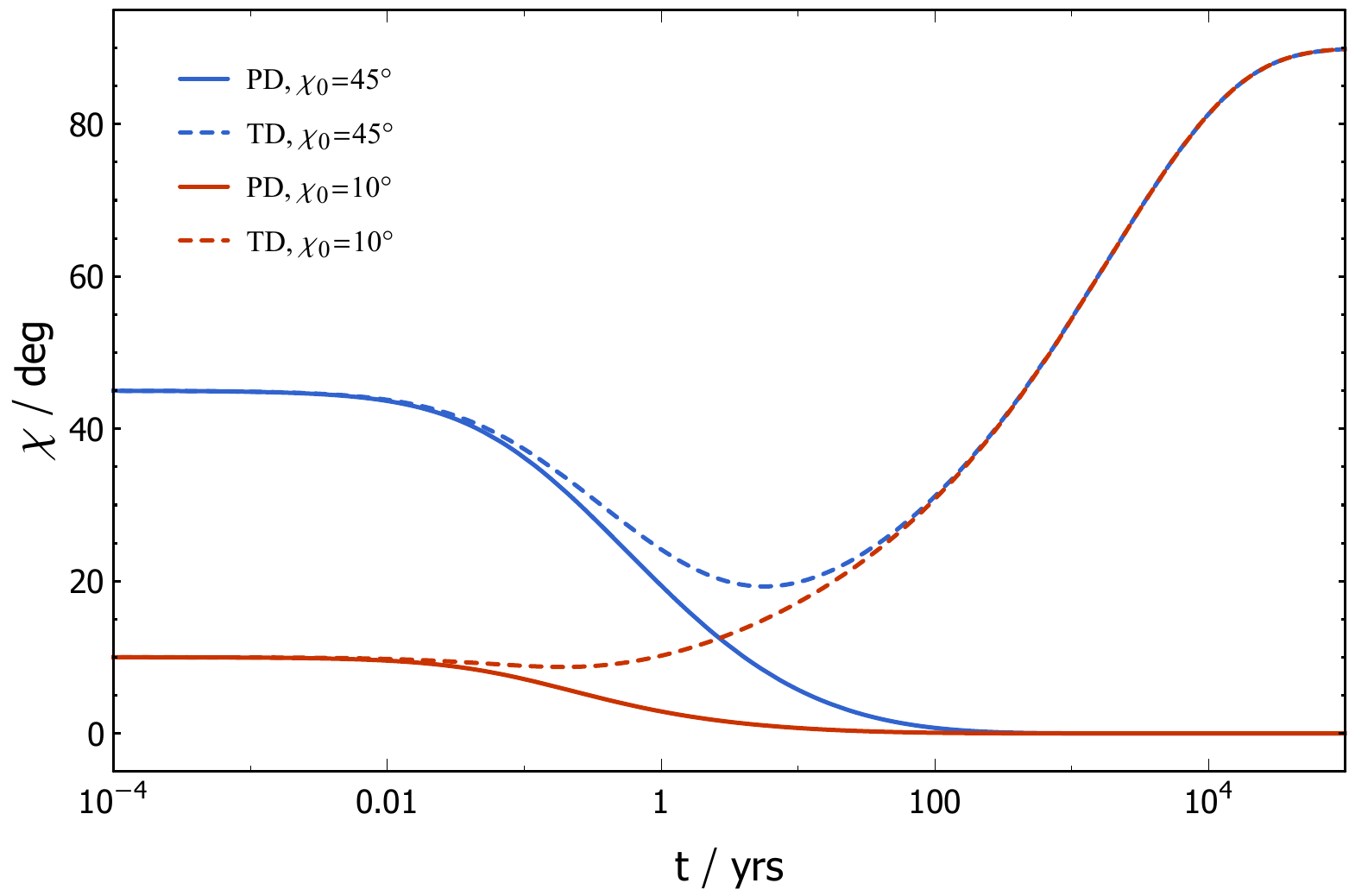}
\caption{Time-dependent evolution of the magnetic inclination angle $\chi$ for Swift J1834. Dashed lines: TD case, evolving toward orthogonality ($\chi \rightarrow 90^\circ$). Solid lines: PD case, trending toward alignment 
 ($\chi \rightarrow 0^\circ$). Colors indicate initial values: red for $\chi_0 = 45^\circ$, blue for $\chi_0 = 10^\circ$.}
\label{fig:chi_t}
\end{figure}

The evolution of the magnetic inclination angle $\chi(t)$ is shown in Figure\,\ref{fig:chi_t}. In the TD case (dashed lines), the inclination angle exhibits non-monotonic behavior: for small $\chi_0$ (red dashed curve), $\chi$ first increases due to the dominance of viscous dissipation (driven by the $\epsilon_\mathrm{B} < 0$ prolate deformation), then stabilizes and approaches an orthogonal configuration ($\chi \rightarrow 90^\circ$). For medium $\chi_0$ (blue dashed curve), $\chi$ initially decreases slightly due to magnetic dipole radiation torque, but eventually evolves toward orthogonality as viscous damping take over. This behavior reflects the competition between terms in Equation\,(\ref{eq:dchi}): the first two terms (MDR and GWE) tend to alignment configuration ($\chi \rightarrow 0^\circ$), while the third term (viscous damping) drives the system toward orthogonality for TD case. In contrast, the PD case (solid lines) shows a monotonic decrease in $\chi$, asymptotically approaching alignment, as the oblate deformation ($\epsilon_\mathrm{B} > 0$) causes viscous damping to reinforce the MDR and GWE alignment torques.

\textbf{The convergence of the rotational tracks at late times (Figure\,\ref{fig:omega_t}) indicates that the present-day spin state is largely independent of the initial inclination angle $\chi_0$, as the strong braking over the magnetar's lifetime erases the memory of the initial geometry. The more pronounced divergence at early epochs, however, highlights the sensitivity of the youthful spin-down to the birth geometry. Furthermore, the evolutionary paths of $\chi(t)$ (Figure\,\ref{fig:chi_t}) are not  merely consequences but active drivers of the spin-down. The progression toward an orthogonal rotator ($\chi \to 90^\circ$) in the TD case, for instance, enhances the braking torque, creating a feedback loop that accelerates rotational energy loss. These coupled evolutionary tracks demonstrate that the present-day braking index $n \approx 1$ is a natural outcome of the system's long-term evolution within our model, rather than a transient state.}

A deeper exploration of the parameter space reveals that the inclination evolution is highly sensitive to 
the precession cycle parameter $\xi$ and the wind parameter $\kappa$. When $\xi$ is small ($\sim 10^4$), viscous dissipation dominates, causing $\chi$ to increase rapidly toward orthogonality; when $\xi$ is large ($\sim 10^6$), radiative braking\,(MDR and GWE) dominates, leading to a decrease in $\chi$ toward alignment. This sensitivity underscores the importance of constraining $\xi$ through multi-messenger observations, as discussed in Section \ref{sec:constration_xi}. Additionally, we find that magnetic field decay timescale and dipole strength exert minimal influence on long-term evolutionary trajectories, as their effects are secondary to the wind braking and viscous dissipation mechanisms. However, angular velocity evolution is highly sensitive to $\kappa$: larger $\kappa$ values enhance wind-dominated energy loss, accelerating spin-down and indirectly slowing the evolution of $\chi$ by reducing the rotational energy available for torque adjustments.

These results have important implications for the evolutionary classification of magnetars. The TD configuration drives Swift J1834 toward a state of orthogonal rotation ($\chi \approx 90^\circ$), which maximizes continuous gravitational wave emission. In contrast, the PD configuration leads to aligned rotation ($\chi \approx 0^\circ$), reducing GW emission but potentially increasing electromagnetic visibility through pulsed radiation. The fact that both configurations can reproduce the current observed state of Swift J1834 suggests 
that additional constraints (e.g., from X-ray polarization or nebular properties) are needed to break the degeneracy. 

\subsection{Inference of Initial Parameters}
\label{sec:Inference of Initial Parameters}
To reconstruct the evolutionary history of Swift J1834 from birth to the present epoch, we perform Bayesian inference of its key initial parameters based on the currently observed quantities, including the spin period $P_\mathrm{obs}$, period derivative $\dot{P}_\mathrm{obs}$, and present magnetic field strength $B_{\mathrm{d}}$. The torque model is the same as in Section~\ref{sec: model}. \textbf{Additionally, we consider the evolution of parameter $\kappa$ in the form of 
$\kappa(t) = \kappa_0 (B(t) / B_0)^{\beta}$, where $\kappa_0$ is a normalization factor and $\beta$ parametrizes the effective sensitivity of the pair multiplicity to the field strength. In our torque model, $\kappa(t)$ enters only through the wind term. To incorporate constraints from the X-ray wind nebula, we impose an additional Gaussian likelihood on the present-day value $\kappa_{\rm now}\equiv\kappa(t_\mathrm{age})$, centred on the range $\kappa_{\rm now}\simeq 13$--$37$ inferred from the nebular efficiency (Section~\ref{sec:Wind Braking}).}

The sampling parameters are $\theta = (P_0, \chi, \kappa_0, \beta, t)$, while $\tau_{\mathrm{Ohm}}$ and $\tau_{\mathrm{Hall}}$ are kept fixed as described above. Since there is no direct observational measurement of the inclination angle for Swift J1834, we do not include inclination evolution in the Bayesian inference. Based on the observed magnetic inclination angles of other magnetars (see Table \ref{tab:1}), we adopt a prior of $\cos\chi \sim \mathrm{Uniform}[\cos 10^\circ,\ \cos 70^\circ]$. 
\textbf{$\kappa_0$ is positive and spans several orders of magnitude. However, due to the lack of a consistent scale in the literature, we adopt a broad log-uniform prior with $\log \kappa_0 \sim \mathrm{Uniform} [0, 5]$. For the exponent we take a conservative flat prior $\beta\sim\mathrm{Uniform}[-3,3]$. }

A Jacobian term is included in the log-space sampling to ensure uniformity in the linear timescale. \textbf{The age is treated as a free parameter with a broad, non-informative prior, $\log t\sim\mathrm{Uniform}\,[2,6],$ corresponding to $t\simeq10^{2}$--$10^{6}$\,yrs. This choice allows the timing data and torque model to determine the preferred age without imposing the supernova remnant estimate as a hard prior. }

The rotational evolution is modeled using the torque formulation in Section\,\ref{sec: model}, which includes contributions from magnetic dipole radiation, gravitational wave emission, and particle wind as given by Equation\,(\ref{eq:domega}), neglecting changes in the stellar mass or moment of inertia. A widely discussed scenario for the origin of magnetars suggests that they are born from proto-neutron stars with initial spin periods $P_0 \sim 0.6-3\,\mathrm{ms}$. The convection and differential rotation can drive an efficient $\alpha$-dynamo mechanism, amplifying the magnetic field to magnetar strength.\citep{Duncan1992ApJ, Thompson1993ApJ}.  

Three main approaches have been proposed to constrain magnetar birth periods. The first is to model supernova light curves or GRB afterglow plateaus with a nascent magnetar spin-down injection model, thereby inferring the initial spin and magnetic field; this generally yields millisecond-scale periods\,\citep{Kasen2010ApJ,Bernardini2015JHEAp}. The second is to compare the explosion energies of SNRs associated with magnetars against the canonical supernova energy ($\sim10^{51}$\,erg). If $P_0 \sim 1$\,ms, the rotational energy released could exceed the canonical value by an order of magnitude. However, \citet{Vink2006MNRAS} showed that several well-studied examples (Kes 73, CTB 109, N49) exhibit standard energies near $10^{51}$ erg, implying $P_0 \gtrsim 5$ ms. \citet{Yan2024EPJC} combined SNR energy upper limits for slowly rotating magnetars with evolutionary models, showing that in weak supernovae the minimum period must be $\sim 5-6$\,ms assuming equality between the rotational energy and the ejecta kinetic energy. The third method is population synthesis. By performing magnetar population synthesis with Monte Carlo simulations and optimization against observed samples, \citet{Jawor2022MNRAS} derived a loose upper limit $P_0<2$\,s. By using Monte Carlo population synthesis with spin, inclination evolution, and magnetic-field decay in vacuum/plasma magnetospheres, \citet{Huang2024PhRvD} obtained expected initial periods in the range $0.03-3.45$\,s.  

\textbf{These discrepancies likely reflect the diversity of magnetar formation channels. For the initial spin period $P_0$ we consider two complementary priors. Motivated by fast-born magnetar scenarios, one choice is a log-normal prior. $\log_{10}P_0 \sim \mathcal{N}(-2.3,0.5),$ which favours $P_0$ of a few milliseconds but still allows for order-of-magnitude variations. To assess how strongly the data constrain $P_0$, we also perform a run with a broad log-uniform prior on $P_0$ over the same range. 
For each parameter sample $\theta$ we integrate the spin evolution from $t=0$ to $t_\mathrm{age}$ on a geometric time grid, obtaining $P(t)$ and $\dot{P}(t)$ as well as the derived initial values $\dot{P}_0$ and $B_0$. To reduce dynamic-range effects we use a Gaussian likelihood in log space,}

\begin{equation}
\ln \mathcal{L}(\theta)
=
-\frac{1}{2}
\left[
  \frac{\log P(\theta) - \log P_{\mathrm{obs}}}{\sigma_{\log P}}
\right]^2
-\frac{1}{2}
\left[
  \frac{\log \dot P(\theta) - \log \dot P_{\mathrm{obs}}}{\sigma_{\log \dot P}}
\right]^2
-\frac{1}{2}
\left[
  \frac{\log \kappa_{\mathrm{now}}(\theta) - \mu_{\log \kappa_{\mathrm{now}}}}{\sigma_{\log \kappa_{\mathrm{now}}}}
\right]^2 ,
\label{eq:likelood}
\end{equation}
where $\sigma$ terms denote uncertainties. Sampling is performed with the \texttt{emcee} ensemble sampler \citep{Foreman2013ascl}, and the posterior structure and parameter degeneracies are visualized with corner plots.

\begin{figure}[h]
\centering
\begin{minipage}{0.48\linewidth}
\centering
\includegraphics[width=\linewidth]{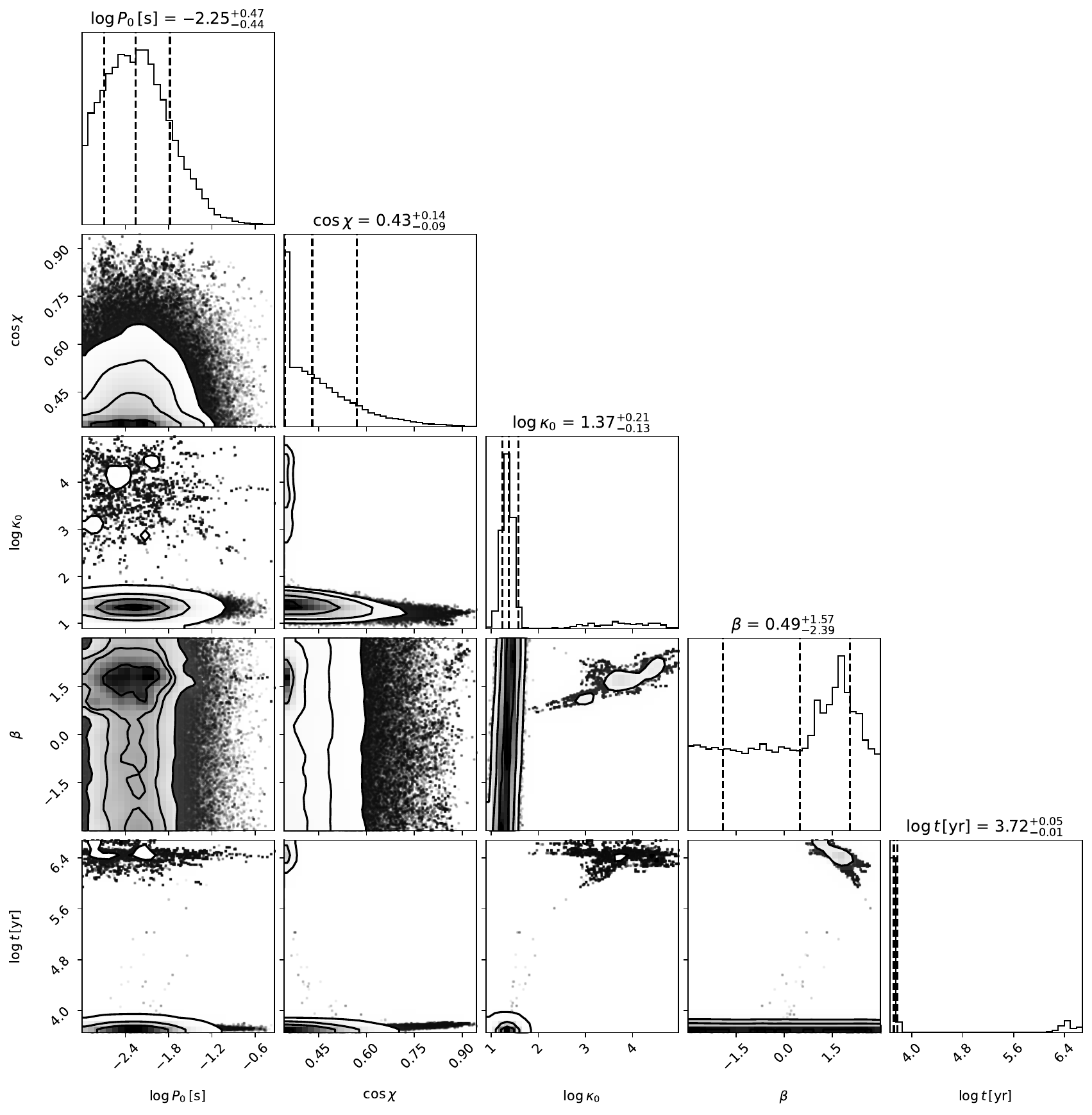}
\end{minipage}\hfill
\begin{minipage}{0.48\linewidth}
\centering
\includegraphics[width=\linewidth]{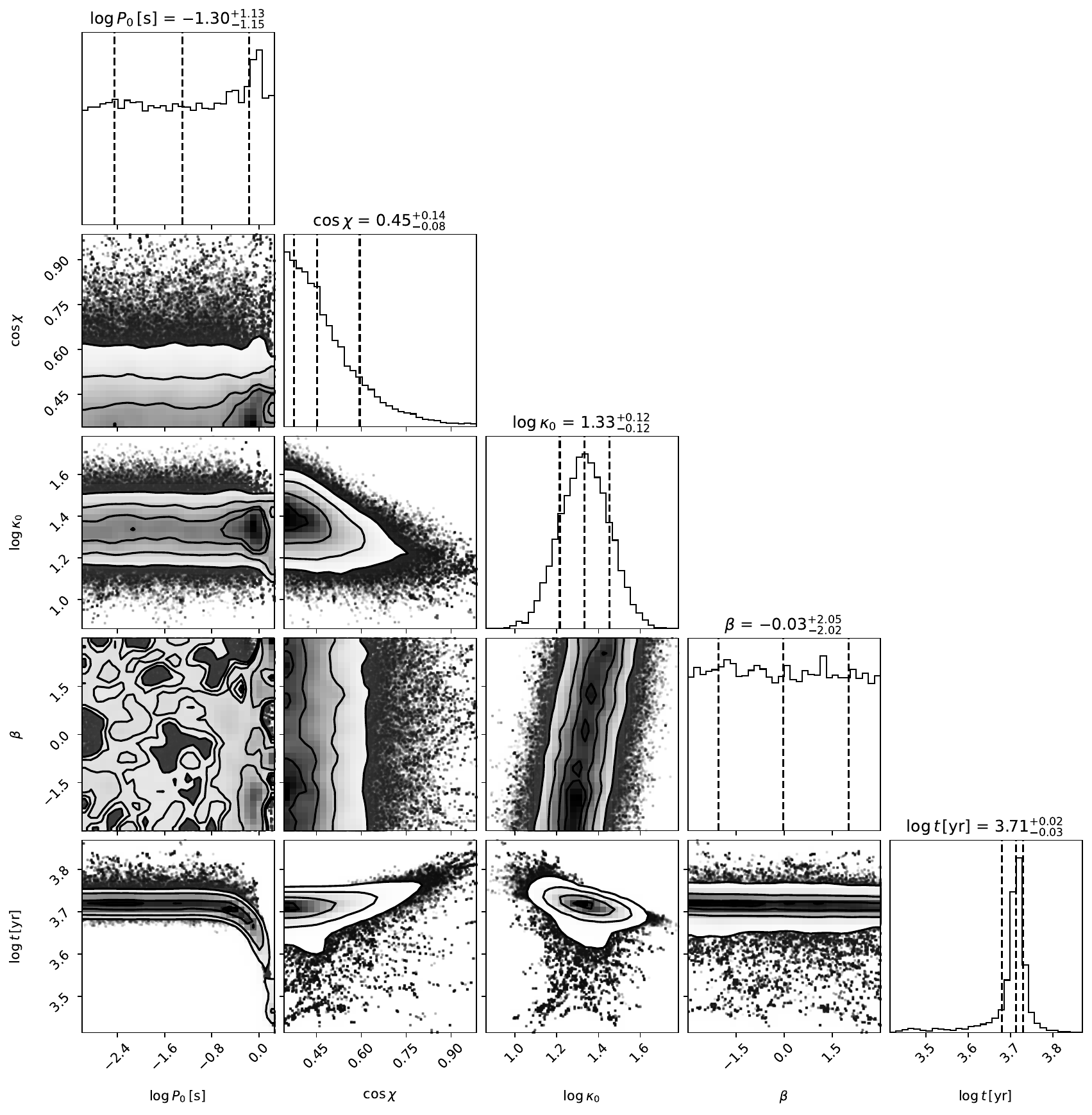}
\end{minipage}
\caption{Posterior distributions of the inferred parameters for Swift J1834 under two different priors on the initial spin period $P_0$. Left: log-normal prior on $P_0$; right: log-uniform prior on $P_0$. The diagonal panels show the marginalized one-dimensional posteriors with the median and 68\% credible intervals, while the off-diagonal panels display the two-dimensional joint posteriors.}
\label{fig:corner_plot}
\end{figure}

\textbf{Figure\,\ref{fig:corner_plot} shows the posterior of Bayesian inference. The inclination angle is constrained to $\chi \simeq 54^\circ$--$70^\circ$, and $\kappa_0 \approx 23$. The exponent $\beta$ remains only weakly constrained: the posterior is broad and centred near zero, with a $68\%$ credible interval approximately $-2\lesssim\beta\lesssim 2$. The detailed time dependence of $\kappa$ should therefore be regarded as a source of systematic uncertainty rather than a robust measurement.
The age is tightly constrained by the timing data within this evolution model, with a posterior peak at $t\simeq 6\times10^3$\,yrs and a narrow credible interval of a few per cent. The initial dipole field is only slightly larger than the present dipole field, $B_0\simeq1.5\times10^{14}$\,G, and the corresponding initial period derivative is of order $\dot{P}_0 \sim 10^{-9}\,\mathrm{s}\,\mathrm{s}^{-1}$, reflecting strong early spin-down. Evolving posterior samples forward within the adopted model yields present-day periods and period derivatives $P_{\mathrm{pred}}$ and $\dot{P}_{\mathrm{pred}}$ that agree with $P_\mathrm{obs}$ and $\dot{P}_\mathrm{obs}$, confirming the internal consistency of the Bayesian framework.
The behaviour of the initial period $P_0$ deserves special comment. Under the log-normal prior the posterior peaks at a few milliseconds, which is compatible with convective-dynamo expectations. However, when the broad log-uniform prior is used, the posterior for $\log_{10}P_0$ becomes nearly flat across the prior range, indicating that the current timing and nebular data provide only weak direct information on $P_0$. The inference of a millisecond birth period is therefore strongly prior-dependent: the data are consistent with a fast-born magnetar, but do not by themselves require it. }

\begin{figure}[h]
    \centering
    \includegraphics[width=0.7\linewidth]{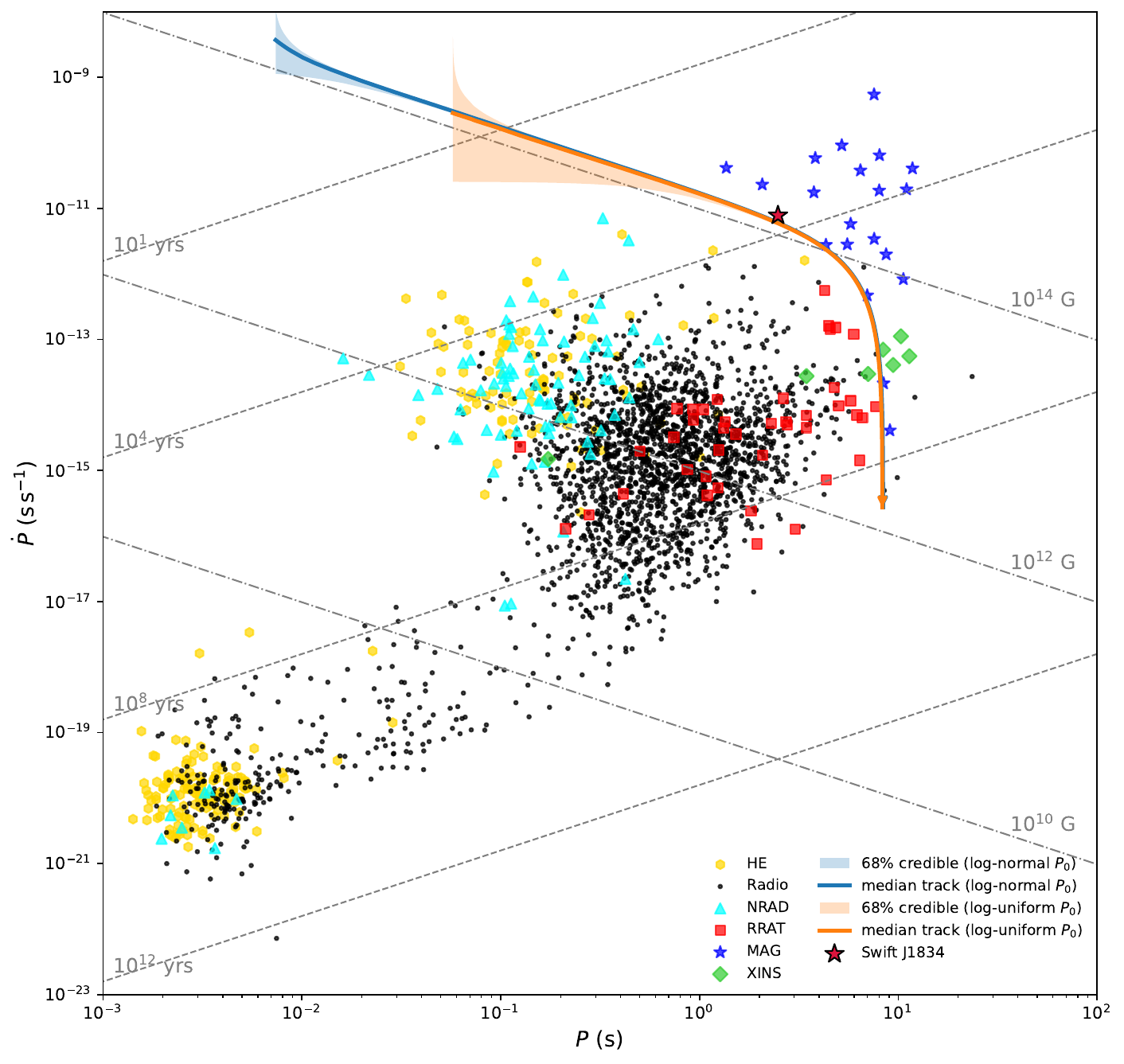}
    \caption{Evolution tracks of Swift J1834 in the $P$--$\dot{P}$ diagram, derived from Bayesian posterior samples. The blue (orange) curve shows the median track obtained with a log-normal (log-uniform) prior on the initial period $P_0$, and the corresponding shaded regions indicate the pointwise
    $68\%$ credible intervals. The red star marks the current observed position of Swift J1834. Different markers indicate various neutron star populations: high-energy pulsars (HE, yellow circles), normal radio pulsars (black dots), non-radio active pulsars (NRAD, cyan triangles), rotating radio transients (RRATs, red squares), X-ray isolated neutron stars (XINSs, green diamonds), and magnetars (MAG, blue stars). These data are taken from \citet{Manchester2005AJ} and \citet{Olausen2014ApJS}.}
    \label{fig:p_pdot}
\end{figure}

Figure\,\ref{fig:p_pdot} shows the $P$--$\dot{P}$ evolutionary track of Swift J1834, derived from the posterior samples obtained through Bayesian analysis. The blue and orange curves represent the median tracks for the runs with a log-normal and a log-uniform prior on the initial period $P_0$, respectively, while the corresponding shaded regions indicate the pointwise $68\%$ credible bands.  The red star marks the current observed position of Swift J1834. In both cases, the median track passes through the observed data point, demonstrating that the torque model can
reproduce the present-day spin state in a self-consistent manner.

\textbf{The overall evolutionary trend is similar for the two priors. Starting from a rapidly spinning, strongly magnetized state, Swift J1834 moves toward longer periods and much lower $\dot{P}$ as the dipole field decays and the star spins down, eventually approaching the region populated by XINS-like objects. 
In the log-normal case the early-time 68\% credible band is concentrated around millisecond periods, whereas in the log-uniform run the corresponding shaded region is noticeably broader in $P$, reflecting the wider range of birth periods allowed by the prior. Because the early spin-down is extremely rapid, the two median tracks quickly converge once $P\gtrsim 0.02\,$s, and the parts of the trajectories relevant for the current state of Swift J1834 are almost indistinguishable. The limited number of observables and the lack of direct constraints on parameters such as $\kappa_0$ and $\chi$ both contribute to this prior sensitivity.
Future improvements could come from incorporating additional observables, such as X-ray polarization measurements to constrain $\chi$, long-term monitoring of $\dot{P}$ and glitches, and X-ray energetics of the wind nebula to better inform $\kappa(t)$. A hierarchical prior on $P_0$ informed by the population distribution of magnetars and high-$B$ pulsars would also provide a more data-driven way to encode birth-spin information.}

\textbf{It is worth noting that, when no SNR-based prior is imposed on the age, both of our Bayesian analyses yield $t \simeq 6\times 10^{3}\,\mathrm{yrs}$, which differs by more than an order of magnitude from the age estimate of W41. Within the framework of the current evolutionary model, we are unable to find any set of parameters that can simultaneously reproduce $P$, $\dot{P}$ and $t_{\mathrm{snr}}$. If we nevertheless assume that the association with W41 is real, then reconciling the age inferred from our evolution model with the SNR age would require a much more complex long-term torque evolution history than we have considered here. Long-term X-ray monitoring shows bursts and transient behaviour from Swift J1834.9$-$0846, but no clear glitch or anti-glitch has been reported. It is therefore more reasonable to attribute any additional torque to extended phases of enhanced wind power associated with bursts/activity, to magnetospheric reconfiguration, or to more complex magnetic-field evolution, rather than to frequent, large-amplitude glitches. Of course, glitches occurring during the early evolutionary phase cannot be ruled out, but current observations do not constrain such early events. A self-consistent treatment of such non-stationary torques would require introducing piecewise braking indices or burst-driven torque terms into the present model, which is beyond the scope of this work. Our inferred age and initial parameters should therefore be interpreted as conditional on the assumption of a smooth, secular spin-down history without major torque excursions.}

\subsection{Constraints on the Number of Precession Cycles $\xi$ and Internal Magnetic Field Structure}
\label{sec:constration_xi}

This section presents a comprehensive analysis of the constraints on the precession cycle parameter $\xi$ and the internal magnetic field configuration of Swift J1834, based on the coupling between the braking index and magnetic inclination evolution.
\textbf{Here we describe the dipole-field evolution in terms of a single effective decay timescale $\tau_\mathrm{d}$, defined by
$\dot B_\mathrm{d} \equiv -B_\mathrm{d}/\tau_\mathrm{d}$ at the present epoch.}
The parameter $\xi$, defined as the ratio of the viscous dissipation timescale to the precession period ($\xi \equiv \tau_{\text{DIS}} / P_{\text{prec}}$), quantifies the efficiency of internal damping processes and plays a crucial role in determining the evolutionary trajectory of the magnetic inclination angle $\chi$. By combining observational constraints from the braking index $n = 1.08 \pm 0.04$ with theoretical models of magnetic field decay, we derive allowed ranges for $\xi$ and use these to discriminate between TD and PD internal field geometries.

Guided by the low braking index and the presence of a bright wind nebula (Section~\ref{sec:omega_chi_Evolution}), we first consider a TD configuration as a reference case. To constrain $\xi$, we substitute the observed values of the spin period $P = 2.48~\text{s}$, period derivative $\dot{P} = 0.796 \times 10^{-11}~\text{s}~\text{s}^{-1}$, and braking index $n = 1.08$ into the braking index expression (Equation\,(\ref{eq:n})) and the magnetic inclination evolution equation (Equation\,(\ref{eq:dchi})). Treating $\xi$ as a free parameter and fixing the wind parameter to a representative value $\kappa = 25$, we solve the coupled system numerically to obtain the relationship between the magnetic field decay timescale $\tau_\mathrm{d}$ and the inclination angle $\chi$ for different values of 
$\xi$.

\begin{figure}[htbp]
\centering
\includegraphics[width=0.7\textwidth]{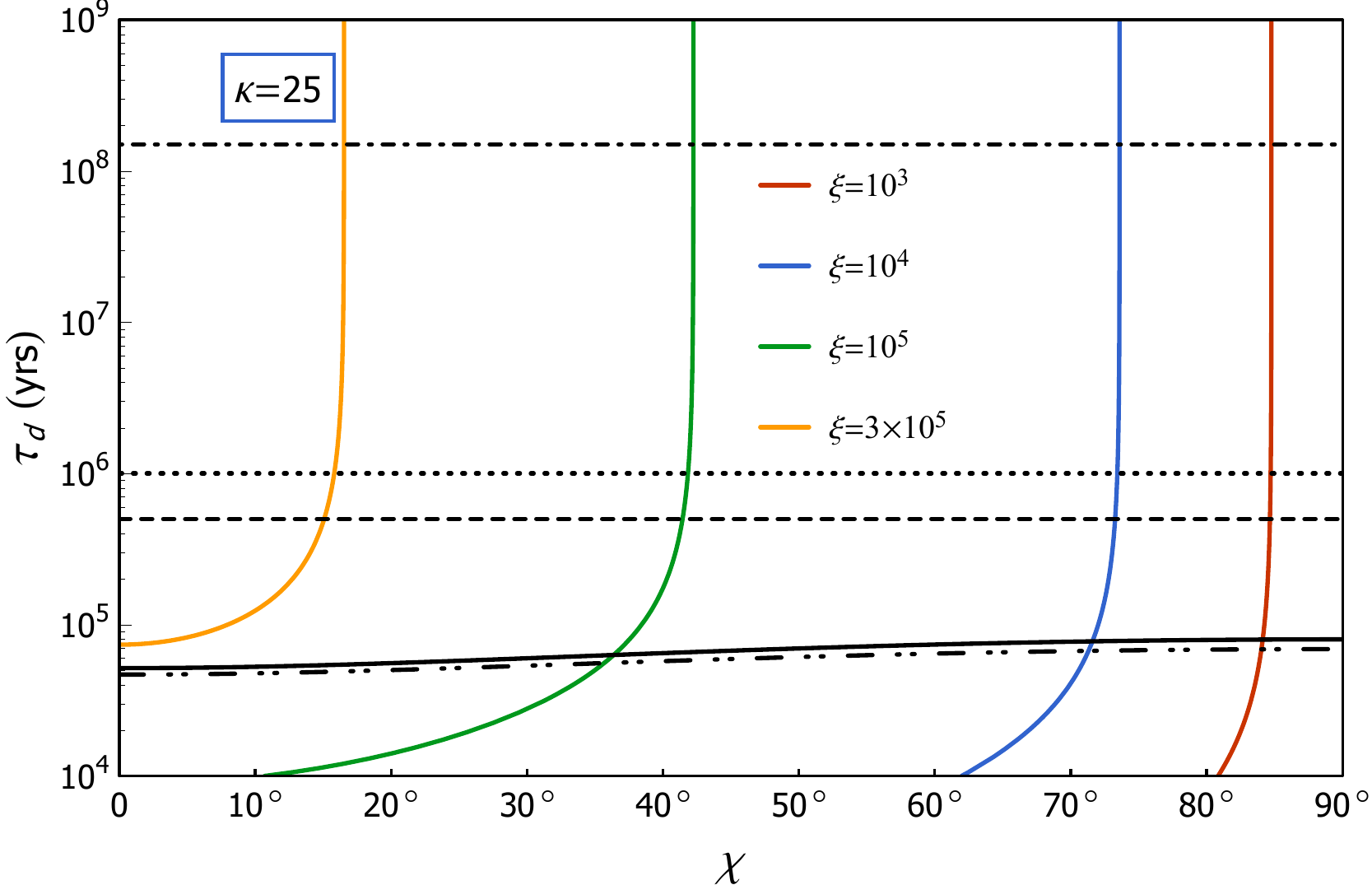}
\caption{Relationship between the dipole magnetic field decay timescale $\tau_\mathrm{d}$ and the magnetic inclination angle $\chi$ for Swift J1834. Colored curves represent model predictions for different values of the precession cycle parameter $\xi$ (legend). Black lines show theoretical decay timescales: Ohmic dissipation (dash-dotted and dashed), Hall drift (solid), and effective minimum (dash-dot-dotted). Intersections define constraints on $\xi$.}
\label{fig:td_chi}
\end{figure}

Figure\,\ref{fig:td_chi} shows the resulting $\tau_\mathrm{d}$--$\chi$ curves for $\xi$ ranging from $10^3$ to $10^5$, along with theoretical decay timescales from magnetic field evolution models. In Figure\,\ref{fig:td_chi}, the black dash-dotted, dotted, and dashed lines show the Ohmic timescales for magnetic-field decay dominated by Ohmic dissipation. From the timing data of Swift J1834, the magnetic-field strength can be expressed as a function of $\chi$, and thus the Hall timescale $\tau_{\mathrm{Hall}}$ also depends on $\chi$ (black solid line). Adopting $\tau_{\mathrm{Ohm}} = 5\times10^5$\,yrs yields a minimum effective timescale $\tau_{\mathrm{eff}}$, which likewise varies with $\chi$ (black dash-dot-dotted line). The intersection points between the colored curves (model-derived $\tau_\mathrm{d}$) and black lines (theoretical timescales) define the allowed values of $\xi$ at each $\chi$.

As shown in Figure\,\ref{fig:td_chi}, the model-derived $\tau_\mathrm{d}$--$\chi$ curves intersect with the theoretical decay timescales across the entire range of inclination angles ($10^\circ \leq \chi \leq 70^\circ$). This indicates that $\xi$ can be constrained for any plausible value of $\chi$. We focus on the current inclination angle $\chi = 45^\circ$ as a representative case, consistent with magnetar population studies (Table\,\ref{tab:1}). Following the $\kappa$--$\chi$ relation from Section\,\ref{sec:Wind Braking}, we consider three values of the wind parameter: $\kappa = 13, 25, 37$, spanning the observed range for Swift J1834. For each $\kappa$, we substitute $\chi = 45^\circ$ into Equations\,(\ref{eq:n}) and (\ref{eq:dchi}) to express $\tau_\mathrm{d}$ as a function of $\xi$. By equating this function with the minimum and maximum theoretical decay timescales (dot-dashed and dash-dot-dotted lines in Figure\,\ref{fig:td_chi}, we solve for the lower and upper limits of $\xi$, denoted $\xi_{\min}$ and $\xi_{\max}$.

The constraints on $\xi$ are summarized in Table\,\ref{tab:3}. The results indicate that $\xi$ lies in the range $\sim 1.1 \times 10^4$ to $3.0 \times 10^5$, with the exact value depending on $\chi$ and $\kappa$. For $\chi = 45^\circ$ and $\kappa = 25$, the constrained range is $\xi = 6.47 \times 10^4$ to $8.68 \times 10^4$. This range is consistent with theoretical estimates for young neutron stars $\xi \sim 10^3$ to $10^6$ \citep{Hu2023RAA, Cheng2019PhRvD}, and reflects the efficiency of viscous damping in the stellar interior. The dependence on $\chi$ is monotonic: higher inclination angles yield smaller 
$\xi$ values, as orthogonal configurations enhance gravitational wave emission and reduce the damping timescale.

\begin{table}[h]
    \caption{Constraints on the precession cycle parameter $\xi$ for different values of the magnetic inclination angle $\chi$ and wind parameter $\kappa$.}
    \label{tab:3}
    \centering
    \renewcommand{\arraystretch}{1.6}
    \begin{tabular}{ccccccc}
    \toprule
    \multirow{2}{*}{$\chi$}  
    & \multicolumn{2}{c}{$\kappa = 13$} 
    & \multicolumn{2}{c}{$\kappa = 25$} 
    & \multicolumn{2}{c}{$\kappa = 37$} \\  
    \cmidrule(l){2-3} \cmidrule(l){4-5} \cmidrule(l){6-7}
    & $\xi_{\mathrm{min}}$  & $\xi_{\mathrm{max}}$  
    & $\xi_{\mathrm{min}}$  & $\xi_{\mathrm{max}}$  
    & $\xi_{\mathrm{min}}$  & $\xi_{\mathrm{max}}$  \\
    \midrule
    $10^{\circ}$  & $2.30\times10^{5}$ & $3.13\times10^{5}$ & $2.35\times10^{5}$  & $3.62\times10^{5}$ & $2.40\times10^{5}$ & $4.30\times10^{5}$ \\
    $20^{\circ}$  & $1.77\times10^{5}$ & $2.33\times10^{5}$ & $1.81\times10^{5}$ & $2.65\times10^{5}$ & $1.86\times10^{5}$ & $3.08\times10^{5}$ \\
    $30^{\circ}$  & $1.23\times10^{5}$ & $1.57\times10^{5}$ & $1.26\times10^{5}$ & $1.76\times10^{5}$ & $1.30\times10^{5}$ & $2.01\times10^{5}$ \\
    $40^{\circ}$  & $8.01\times10^{4}$ & $1.00\times10^{5}$ & $8.24\times10^{4}$ & $1.11\times10^{5}$ & $8.48\times10^{4}$ & $1.25\times10^{5}$ \\
    $45^{\circ}$  & $6.29\times10^{4}$ & $7.82\times10^{4}$ & $6.47\times10^{4}$ & $8.68\times10^{4}$ & $6.67\times10^{4}$ & $9.76\times10^{4}$ \\
    $50^{\circ}$  & $4.83\times10^{4}$ & $5.96\times10^{4}$ & $4.97\times10^{4}$ & $6.61\times10^{4}$ & $5.13\times10^{4}$ & $7.42\times10^{4}$ \\
    $60^{\circ}$  & $2.59\times10^{4}$ & $3.17\times10^{4}$ & $2.67\times10^{4}$ & $3.51\times10^{4}$ & $2.76\times10^{4}$ & $3.92\times10^{4}$ \\
    $70^{\circ}$  & $1.11\times10^{4}$ & $1.35\times10^{4}$ & $1.15\times10^{4}$ & $1.50\times10^{4}$ & $1.19\times10^{4}$ & $1.67\times10^{4}$ \\
    \bottomrule
    \end{tabular}
\end{table}

\begin{figure}[h]
\centering
\includegraphics[width=0.7\linewidth]{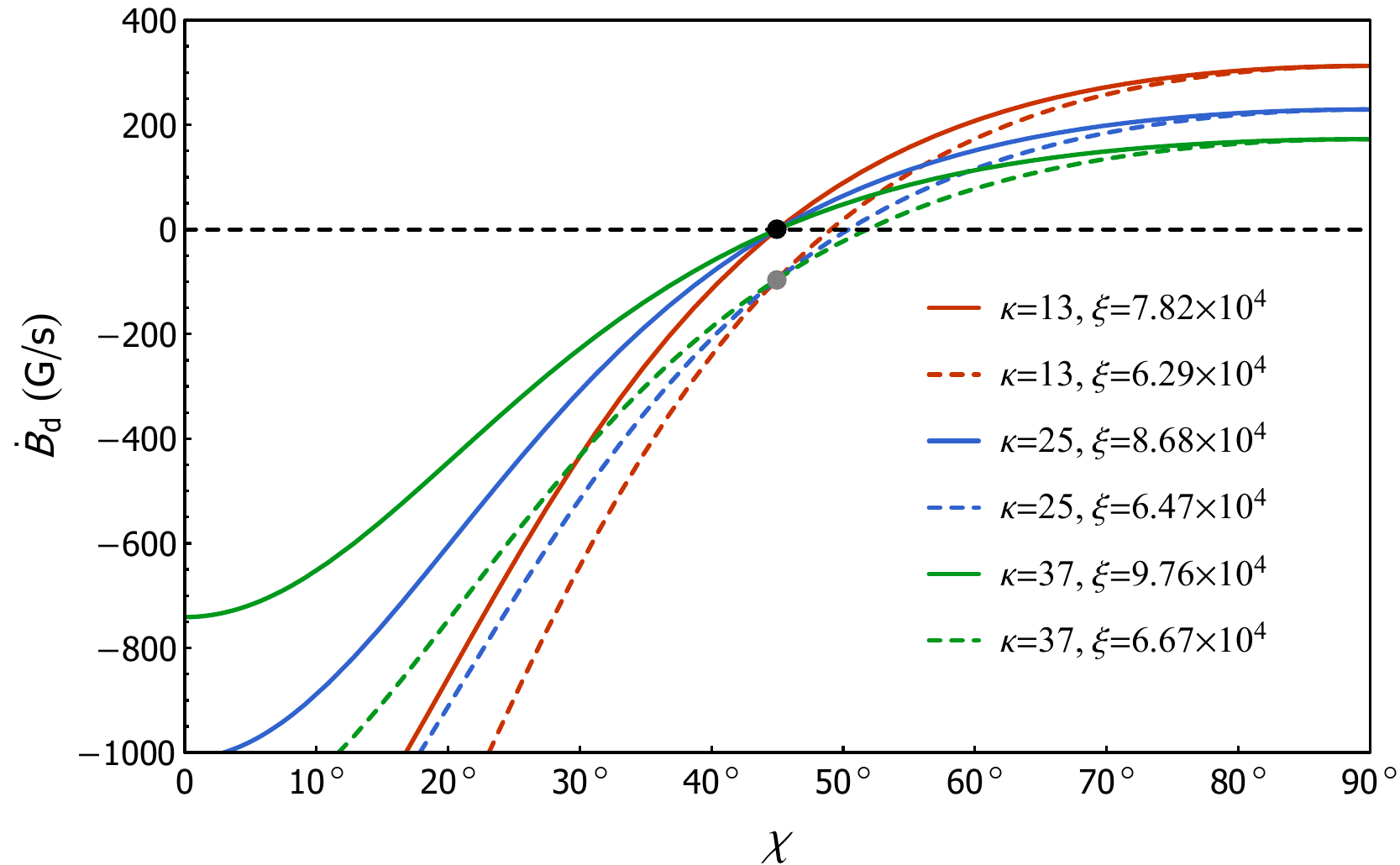}
\caption{Relationship between the magnetic field decay rate $\dot{B}_\mathrm{d}$ and the inclination angle $\chi$ for Swift J1834 under the TD configuration. Solid lines: $\xi = \xi_{\max}$; dashed lines: $\xi = \xi_{\min}$. Colors denote $\kappa$ values: red ($\kappa = 13$), green ($\kappa = 25$), blue ($\kappa = 37$). The black and gray dots mark $\dot{B}_\mathrm{d}$ at $\chi = 45^\circ$ for $\xi_{\max}$ and $\xi_{\min}$, respectively.}
\label{fig:db_chi}
\end{figure}

To further validate the TD configuration, we examine the implications for magnetic field decay. 
Substituting the constrained $\xi$ values back into Equations\,(\ref{eq:n}) and (\ref{eq:dchi}) for $\chi = 45^\circ$, we compute the corresponding magnetic field decay rate $\dot{B}_\mathrm{d}$ as a function of $\chi$ for $\kappa = 13, 25, 37$. Figure\,\ref{fig:db_chi} shows the resulting $\dot{B}_\mathrm{d}$--$\chi$ relations. In the TD case, $\dot{B}_\mathrm{d} < 0$ for all $\chi$, consistent with the physical requirement of magnetic field decay. For $\xi = \xi_{\max}$ (solid lines), $\dot{B}_d$ is 
relatively slow (e.g., $-0.1~\text{G/s}$ at $\chi = 45^\circ$, black dot), while for $\xi = \xi_{\min}$ (dashed lines), $\dot{B}_\mathrm{d}$ is rapid (e.g., $-95.5~\text{G/s}$ at $\chi = 45^\circ$, gray dot). These values span the plausible range of decay rates for magnetars, supporting the self-consistency of the model.

For comparison, we repeat the above analysis for a PD configuration, in which the internal field produces an oblate deformation ($\epsilon_B > 0$).  Within the same class of monotonic decay laws parametrized by a single effective timescale $\tau_\mathrm{d}$, reproducing the observed low braking index $n < 3$ then requires $\dot{B}_\mathrm{d} > 0$ for all allowed values of $\xi$, i.e.\ a growing dipole field. Such behaviour is difficult to reconcile with standard expectations of field decay in magnetars, and in this phenomenological framework the PD configuration is therefore disfavoured.  We emphasize, however, that this conclusion is model-dependent: more complex,
non-monotonic evolution of $B_\mathrm{d}(t)$ could in principle alleviate this tension, and a fully self-consistent treatment of such scenarios is beyond the scope of the present work.

\textbf{X-ray observations indicate that the persistent emission of Swift~J1834 originates from a small hot region on the neutron-star surface.  Spectral fits with a single blackbody yield temperatures $kT \simeq 0.7$--$1$~keV and an emitting radius $\simeq 0.2$--$0.4$~km, together with an unusually high pulsed fraction $\gtrsim 80\%$ in the 2-10~keV band \citep{Kargaltsev2012ApJ,Esposito2013MNRAS}.  These properties are naturally interpreted as thermal radiation from a small hot spot near a magnetic pole rather than from the entire stellar surface.  Magneto-thermal simulations have shown that strong internal toroidal fields can channel heat and currents into localized surface regions and produce such hot spots and large pulse fractions \citep{Geppert2014MNRAS,Vigano2013MNRAS}.
Combined with the high wind efficiency inferred from the extended nebula (Section~\ref{sec:Wind Braking}), these features are therefore compatible with, and mildly motivate, a TD internal configuration for Swift~J1834.  However, they do not uniquely determine the internal field geometry; other configurations with complex magnetospheric twists or offset dipoles could in principle produce similar surface anisotropies.}

\textbf{The constrained range of the precession parameter, $\xi \sim 10^4$--$10^5$, implies an internal viscous dissipation timescale on the order of $\tau_{\mathrm{DIS}} \sim 10^3$--$10^4$ yrs for a typical magnetar precession period ($P_{\mathrm{prec}} \sim 0.1$ yr). This is consistent with theoretical damping mechanisms such as neutron-star crust-core coupling. Taken together, the hot-spot-like surface emission, the high wind efficiency, and the $\xi$ constraints point towards a scenario where a strong toroidal component plays a major role in Swift J1834. }

\textbf{The consistency of our results exclusively with the TD configuration -- which self-consistently yields a decaying dipole field ($\dot{B}_\mathrm{d} < 0$) and drives the magnetic inclination towards orthogonality -- provides strong evidence for this interpretation. This supports a formation scenario in which a strong toroidal component was generated and sustained, likely by a proto-neutron star dynamo, and continues to govern the magnetar's long-term evolution and energy output. While PD configurations cannot be entirely excluded without more sophisticated models, our analysis robustly identifies the TD scenario as a possible explanation for the observed properties of Swift J1834.
}

\subsection{Synthesis of Results and Implications}
\textbf{The results presented in this section form a coherent narrative that moves from describing the evolutionary history of Swift J1834 to probing its internal physics. The numerical evolution (Section\,\ref{sec:omega_chi_Evolution}) demonstrates the viability of our coupled model in reproducing the current state. The Bayesian inference (Section\,\ref{sec:Inference of Initial Parameters}) quantitatively reverses this process to constrain the probable initial conditions, revealing both the strengths and limitations of the data. Finally, the analysis of the precession dynamics (Section\,\ref{sec:constration_xi}) provides the crucial link between global observables and the hidden internal magnetic field structure. Collectively, these findings robustly support a scenario where wind braking and a TD internal field are key to understanding the anomalous braking index of Swift J1834. This unified picture provides a new framework for interpreting the evolution of magnetars and similar high-B pulsars.}

\section{Gravitational Wave Detectability}
\label{sec: Gravitational Wave}

\subsection{Predicted GW signal from magnetar}
Magnetars are prime targets for continuous GW searches  \,\citep{Cheng2017PhRvD,Yan2024EPJC}. Building on the rotational and magnetic field parameters derived in Section~\ref{sec:Results}, namely the spin frequency, braking index, and the inferred toroidally-dominated internal field configuration, we assess the continuous GW detectability of Swift J1834. These parameters determine both the GW amplitude and the spectral signature, allowing us to evaluate its visibility to current and future detectors In this framework, the emission is fixed by the magnetically induced ellipticity and contains two spectral components\,\citep{Weltevred2011MNRAS,Cheng2015MNRAS}: a fundamental at the stellar spin frequency, $\nu_{e,1}=\nu$, and a second harmonic at twice the spin, $\nu_{e,2}=2\nu$, where $\nu=1/P$ is the spin frequency. The corresponding strain amplitudes are given by the following expressions\,\citep{Wette2023APh}:
\begin{equation}
h_{1}(t) = \frac{8 \pi^{2} G I \epsilon_{B} \nu_{e,1}^{2}}{c^{4} D} \sin(2\chi),
\label{eq:h1}
\end{equation}
and
\begin{equation}
h_{2}(t) = \frac{8 \pi^{2} G I \epsilon_{B} \nu_{e,2}^{2}}{c^{4} D} \sin^{2}\chi,
\label{eq:h2}
\end{equation}
where $D$ is the distance to the source. For Swift J1834, the current spin frequency is $\nu = 0.403$\,Hz (corresponding to a period of $P = 2.48$\,s), and the surface magnetic field is $B_{\mathrm{surf}} = 1.4 \times 10^{14}$\,G. Using Equation\,(\ref{eq:ep_td}), the current magnetically induced ellipticity is calculated to be $\epsilon_\mathrm{B} = -1.4 \times 10^{-6}$. The characteristic strain is computed using the absolute value of $\epsilon_\mathrm{B}$. The characteristic strain amplitudes at frequencies $\nu_{e,1}$ and $\nu_{e,2}$ are given by $h_{c}(\nu_{e,1}) = \nu_{e,1} h_{1}(t)/\sqrt{d\nu_{e,1}/dt}$ and
$h_{c}(\nu_{e,2}) = \nu_{e,2} h_{2}(t)/\sqrt{d\nu_{e,2}/dt}$. 
In our calculations, we adopt the kinematic distance of SNR W41 as the distance to Swift J1834, $D=4.2$\,kpc\,\citep{Leahy2008AJ}. 

\begin{figure}[htb]
    \centering
    \includegraphics[width=0.7\linewidth]{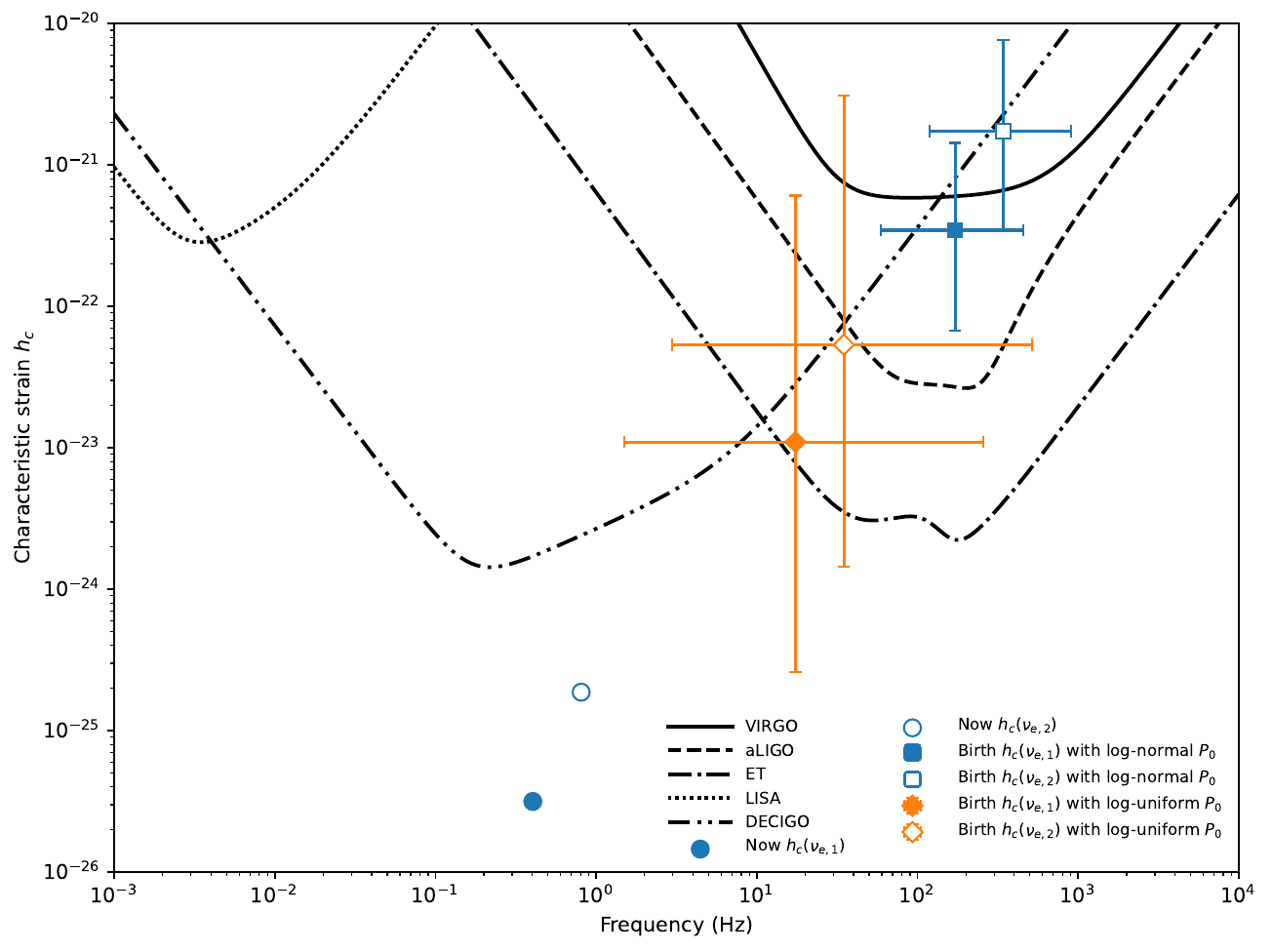}
    \caption{\textbf{Comparison of the sensitivity curves of five detectors with the expected signal strength of Swift J1834. 
    Blue circles mark the present-day fundamental and second-harmonic points. Birth-epoch values propagated from the MCMC posteriors are indicated by symbols with $1\sigma$ error bars: blue squares correspond to the run with a log-normal prior on the initial spin period $P_{0}$, while orange symbols correspond to the run with a broad log-uniform prior on $P_{0}$.
    The curves in the figure indicate the root mean square strain noise levels of the respective detectors\,\citep{Sathyaprakash2009LRR,Yagi2011PhRvD}.}}
    \label{fig:hc_f}
\end{figure}

We calculate $h_c$ of Swift J1834 at birth and at present by propagating the full posterior distributions from our Bayesian analysis (Section\,\ref{sec:Inference of Initial Parameters} through Equations (\ref{eq:h1}) and (\ref{eq:h2}). This approach allows us to quantify the uncertainty in the strain amplitudes arising from the uncertainties in the parameters $\theta$. 
To visualize the gravitational wave signal strength of the magnetar Swift J1834, we compare the computed characteristic strain amplitudes with the root-mean-square (RMS) noise curves of VIRGO, aLIGO, ET, LISA, and DECIGO\,\citep{Sathyaprakash2009LRR,Yagi2011PhRvD}, as shown in Figure\,\ref{fig:hc_f}. 

It can be seen that the present-day points lie far below the sensitivity curves of current and planned detectors, implying that a continuous-wave detection today is unlikely. In contrast, the birth-epoch medians are at the level of $10^{-22}$--$10^{-21}$ near $10^{2}$--$10^{3}$\,Hz, comparable to advanced/third-generation ground-based detectors, indicating potential detectability during a short early phase of rapid spin.

\subsection{Quantitative Detectability Analysis}
\textbf{The analysis presented in Figure\,\ref{fig:hc_f} suggests that while the early-life GW signal from a source like Swift J1834 is a compelling target, its detectability is highly sensitive to the initial conditions, particularly the birth spin period. The significant uncertainty in $P_0$ translates directly into a large uncertainty in the initial GW frequency and strain amplitude. To further quantify the detectability of the birth-epoch signal, we estimate the observing time required for a matched-filter search to reach a given signal-to-noise ratio (S/N). For each of the two Bayesian runs we take the median posterior parameters and compute the S/N for ET as a function of integration time, using the standard definition\citep{Maggioregw}}
\begin{equation}
    (S/N)^2=4\int_0^\infty\frac{|\tilde{h}(f)|^2}{S_n(f)}\mathrm{d}f,
    \label{eq:snr_def}
\end{equation}
\textbf{where $f$ is the gravitational-wave frequency, $\tilde{h}(f)$ is the Fourier transform of the time-domain strain $h(t)$, and $S_{n}(f)$ is the one-sided noise power spectral density of the detector.
Using Equation~(\ref{eq:snr_def}) with the ET design noise curve, we solve for the observing time such that the S/N at the birth epoch satisfies $S/N=8$, for the median posterior parameters of our two inference runs.  
For the run with the log-normal prior on $P_{0}$, the inferred strain is high enough that S/N would reach $8$ after only $3.6\times10^{-2}\,\mathrm{d}$. In contrast, for the run with the broad log-uniform prior on $P_{0}$ the lower characteristic strain implies a much longer integration time: ET would need almost a full year of coherent data to reach the same threshold. }

\textbf{The observing times inferred from Equation\,(\ref{eq:snr_def}) assume an idealized, fully phase-coherent matched-filter search with a perfectly known phase model. This assumption is likely optimistic for a magnetar. Magnetars commonly exhibit strong red timing noise and intermittent glitches/anti-glitches, implying stochastic phase wandering and discrete phase discontinuities that can rapidly accumulate to many radians and invalidate long coherent integrations unless modeled explicitly \citep{Kaspi2017ARAA,Cerri-Serim2019MNRAS,Tsang2013ApJ}. It has been shown that unmodeled timing noise can severely degrade targeted coherent searches, and that glitches can cause substantial mismatch and even missed detections, motivating glitch-robust semi-coherent strategies and tracking approaches \citep{Ashton2015PhRvD,Ashton2018PhRvD,Keitel2019PhRvD,Suvorova2016PhRvD}. Consequently, the physically achievable coherent segment duration $T_{\rm coh}$ for the birth-epoch signal is expected to be much shorter than the values implicitly assumed by the fully coherent estimate. For a fixed total observing time $T_{\rm obs}$, semi-coherent/hierarchical searches incur a sensitivity penalty relative to ideal matched filtering, with strain sensitivity improving only weakly with $T_{\rm coh}$\,\citep{Wette2023APh}. Moreover, the rapid early spin evolution implies large $|\dot{\nu}|$ (and potentially higher derivatives), so the signal frequency can drift appreciably even over short times, further tightening the practical coherent window at the high (up to kHz) birth frequencies. Therefore, our birth-epoch detectability estimates should be regarded as an optimistic upper bound and primarily as an order-of-magnitude benchmark rather than a direct forecast for a realistic search.
}

\textbf{In addition to search-systematics, the inferred birth-epoch S/N is highly sensitive to the poorly constrained initial conditions, particularly the initial spin period $P_0$ and the magnetically induced ellipticity $\epsilon_\mathrm{B}$. As indicated by the posterior distributions in Figure\,\ref{fig:corner_plot}, the data primarily constrain the present-day state, leaving the birth parameters uncertain over a wide range, which maps directly into large variations in the predicted birth frequency and strain amplitude. Correspondingly, the coherent observing time required to reach a given detection threshold can span a broad interval: for example, under a log-uniform prior on $P_0$ the implied $T_{\rm coh}$ for ET can range from $\sim 10^{-2}$\,days to values formally exceeding a year in the most optimistic cases, whereas the physically achievable $T_{\rm coh}$ is likely much shorter in the presence of strong phase wander and glitches. Accordingly, the absolute scale of $h_c$ and the associated observing-time estimates should be interpreted as order-of-magnitude guidance.}

\textbf{Taken together, these caveats imply that, while the birth signal from Swift~J1834 could in principle be detectable by future third-generation detectors under exceptionally favorable conditions, its practical detectability remains highly uncertain. A successful detection would require not only an unusually rapid birth spin and a large ellipticity, but an exceptionally stable early spin evolution that is not strongly affected by the timing irregularities typical of magnetars. More robust constraints on magnetar birth properties are needed before firmer predictions can be made. In any realistic scenario, future searches will likely rely on sophisticated semi-coherent algorithms. Our estimated strain levels are therefore best viewed as an optimistic benchmark for comparison against the sensitivities of practical search pipelines, rather than as a direct forecast of detectability.}

\section{Summary and Discussions}
\label{sec: Conclusion}
This study develops a comprehensive spin-down model incorporating wind braking to investigate the rotational evolution and internal physics of the magnetar Swift J1834. Our work establishes a unified framework connecting three intrinsically coupled physical processes - magnetic field evolution, inclination angle dynamics, and structural deformation. The results demonstrate that in extreme magnetic environments, wind braking contributes significantly to spin-down, challenging the conventional paradigm of magnetic dipole radiation dominance in pulsar braking theory.

\subsection{Summary of Key Findings}
Our numerical modeling and Bayesian inference yield several conclusions about Swift J1834's evolutionary history and internal physics:

\begin{enumerate}
    \item \textbf{Wind Braking Contribution}: \textbf{Wind braking contributes $17\%$--$51\%$ to the total spin-down torque, a result obtained by comparing time-integrated energy loss from wind braking to total rotational energy loss using median posterior parameters from our Bayesian analysis.} This revises standard pulsar spin-down theory by demonstrating that particle outflows play a major role in magnetar rotational evolution.

    \item \textbf{Internal Magnetic Field Configuration}: \textbf{Within our phenomenological framework, the analysis tends to favor a toroidal-dominated (TD) internal magnetic-field structure over a purely poloidal-dominated (PD) configuration.  In the TD case, the observed braking index ($n = 1.08 \pm 0.04$) can be reproduced while maintaining a monotonically decaying dipole field. In contrast, a strictly PD-dominated model with a single monotonic decay timescale would require $\dot{B}_{\mathrm{d}} > 0$ to explain $n<3$, which is difficult to reconcile with the expectation of magnetic-field decay.  We emphasize that this preference for TD is model-dependent: more complex, non-monotonic field evolution could relax the tension in PD scenarios, and the current data do not uniquely fix the internal field topology.}

    \item \textbf{Constraint on the number of precession cycles}: The precession cycle parameter is constrained to $\xi \sim 1.1 \times 10^4 $--$3.0 \times 10^5$, and the inferred range of the precession parameter $\xi$ is compatible with standard neutron star crust-core coupling. This parameter, characterizing viscous dissipation timescales inside the neutron star, provides crucial insight into magnetar interior microphysics and agrees with modern theoretical estimates for young pulsars.

    \item \textbf{Initial parameters of Swift J1834}:
    \textbf{Using Bayesian inference we explore the initial spin under two different priors on the birth period: a log-normal
    prior motivated by millisecond proto-magnetar scenarios and an alternative prior uniform in $\log P_0$ over the same physically allowed range. Under a log-normal prior the posterior peaks at millisecond-scale periods, whereas a broad log-uniform prior yields a nearly flat posterior over the allowed range. }

    \item \textbf{Gravitational-wave emission and detectability:} \textbf{In our posterior samples, gravitational-wave torques contribute only a very
    small fraction of the present-day spin-down power, and the corresponding continuous-wave signal lies well below the sensitivities of current and planned detectors.  At birth, the combination of rapid rotation and strong
    internal fields could have produced characteristic strain amplitudes of
    order $h_c \sim 10^{-23}$--$10^{-21}$, depending on the exact initial spin and ellipticity.  For optimistic corners of parameter space, such an early-time signal would be marginally accessible to advanced LIGO or the Einstein Telescope in the absence of strong timing noise and glitches.}
\end{enumerate}

\subsection{Discussion and Future Outlook}

The wind braking hypothesis finds direct support in Swift J1834's nebular properties, particularly the high conversion efficiency ($\sim 70\%$) of spin-down luminosity to observed nebular radiation. Our model successfully accounts for the observed braking index. Our treatment neglects possible higher-order multipole components that likely exist in real magnetars, which could affect the accuracy of our magnetic field model. Additionally, the axisymmetry assumption, while computationally necessary, could limit applicability to non-axisymmetric emission features, such as those observed in some magnetar 
outbursts. Furthermore, the microphysical link between the wind parameter $\kappa$ and primary particle acceleration mechanisms requires deeper theoretical investigation, as current models are based on simplified gap physics. Although this study focuses on the magnetar Swift J1834.9−0846, the proposed wind-braking framework can be  extended to other low-braking-index sources such as PSR J1734$-$3333 ($n = 0.9 \pm 0.2$) and PSR J1846$-$0258 ($n\approx 0.9$). These objects exhibit detectable X-ray emission and a braking index significantly below 3, suggesting that wind braking and magnetic-inclination evolution may play a common role in their rotational evolution. Future application of this model to a broader sample will test its universality and reveal whether the coupling between wind torque and magnetic-field decay is a generic property of neutron stars with ultra-strong fields.

This work establishes a unified framework linking magnetar spin-down with internal physics and multi-messenger observables, paving the way for several promising research directions. First, applying this methodology to the broader magnetar population may reveal universal relations between magnetic field strength and braking mechanisms, potentially leading to a new classification scheme for neutron stars. \textbf{Second, our study underscores the importance of critically assessing neutron star-SNR associations. An improved and independent age determination for SNR W41 remains highly valuable as a crucial test of our model. A confirmed age of ~130 kyr for W41, if its association with Swift J1834 is upheld, would starkly contradict the ~6 kyr age inferred from our spin-down model and necessitate invoking additional, non-secular evolutionary mechanisms (e.g., early epoch glitches or prolonged outburst activities). Conversely, a much younger age for the SNR would validate our model's timeline and strengthen the association. Thus, resolving this age discrepancy is pivotal to deciphering the true evolutionary history of Swift J1834. Third, testing predicted inclination angle correlations through X-ray polarization observations with missions like  IXPE or eXTP could provide direct evidence for internal field geometries. For Swift J1834, whose evolutionary trajectory favors a present-day inclination in the range of $\chi \simeq 54^\circ$--$70^\circ$ (Section~\ref{sec:Inference of Initial Parameters}), the phase-resolved polarization signature would serve as a decisive probe. A large polarization degree ($> 30\%$) and a sinusoidal modulation of the polarization angle across the pulse phase are expected for such an oblique rotator, clearly distinguishing it from a nearly aligned configuration ($\chi \to 0^\circ$) which would exhibit a low, nearly constant polarization degree. The enhanced sensitivity of next-generation missions will allow a reconstruction of $\chi$ with high precision\,\citep{Ge2025SCPMA}, which can directly test our model's prediction of a toroidal-field-driven evolution toward orthogonality ($\chi \to 90^\circ$).} 
Fourth, conducting targeted searches for continuous gravitational waves from young magnetars using next-generation detectors like the Einstein Telescope is crucial; early-time signals may be detectable, as shown in Figure\,(\ref{fig:hc_f}), offering insights into early magnetar evolution.

Overall, this study provides a new foundation for understanding magnetar evolution by demonstrating the substantial role of wind braking and favoring toroidally dominated internal fields, establishing a broadly applicable research paradigm for connecting magnetar observations with their interior physics.

\section*{Acknowledgements}

This work was supported by the National Natural Science Foundation of China (No.12573052 and No.12573103 ), the National Key Research and Development Program of China (No.2022YFC2205202), the Major Science and Technology Program of Xinjiang Uygur Autonomous Region (No.2022A03013-1), the National Natural Science Foundation of China (No.12288102 and No.12041304), and the Tianshan talents program (2023TSYCTD0013).

\bibliography{ref}{}

@ARTICLE{Olausen2014ApJS,
       author = {{Olausen}, S.~A. and {Kaspi}, V.~M.},
        title = "{The McGill Magnetar Catalog}",
      journal = {\apjs},
         year = 2014,
        month = may,
       volume = {212},
       number = {1},
          eid = {6},
        pages = {6},
          doi = {10.1088/0067-0049/212/1/6},
archivePrefix = {arXiv},
       eprint = {1309.4167},
 primaryClass = {astro-ph.HE},
       adsurl = {https://ui.adsabs.harvard.edu/abs/2014ApJS..212....6O}
}

@ARTICLE{DElia2011GCN,
       author = {{D'Elia}, V. and {Barthelmy}, S.~D. and {Baumgartner}, W.~H. and {Beardmore}, A.~P. and {Chester}, M.~M. and {Guidorzi}, C. and {Holland}, S.~T. and {Kuin}, N.~P.~M. and {Markwardt}, C.~B. and {Marshall}, F.~E. and {Melandri}, A. and {Palmer}, D.~M. and {Saxton}, C.~J. and {Stratta}, G. and {Swenson}, C.~A. and {Ukwatta}, T.~N.},
        title = "{GRB 110807A: Swift detection of a burst or a posible new Sgr.}",
      journal = {GRB Coordinates Network},
         year = 2011,
        month = jan,
       volume = {12253},
        pages = {1},
       adsurl = {https://ui.adsabs.harvard.edu/abs/2011GCN.12253....1D}
}

@ARTICLE{Goldstein2012ApJS,
       author = {{Goldstein}, Adam and {Burgess}, J. Michael and {Preece}, Robert D. and {Briggs}, Michael S. and {Guiriec}, Sylvain and {van der Horst}, Alexander J. and {Connaughton}, Valerie and {Wilson-Hodge}, Colleen A. and {Paciesas}, William S. and {Meegan}, Charles A. and {von Kienlin}, Andreas and {Bhat}, P.~N. and {Bissaldi}, Elisabetta and {Chaplin}, Vandiver and {Diehl}, Roland and {Fishman}, Gerald J. and {Fitzpatrick}, Gerard and {Foley}, Suzanne and {Gibby}, Melissa and {Giles}, Misty and {Greiner}, Jochen and {Gruber}, David and {Kippen}, R. Marc and {Kouveliotou}, Chryssa and {McBreen}, Sheila and {McGlynn}, Sin{\'e}ad and {Rau}, Arne and {Tierney}, Dave},
        title = "{The Fermi GBM Gamma-Ray Burst Spectral Catalog: The First Two Years}",
      journal = {\apjs},
         year = 2012,
        month = mar,
       volume = {199},
       number = {1},
          eid = {19},
        pages = {19},
          doi = {10.1088/0067-0049/199/1/19},
archivePrefix = {arXiv},
       eprint = {1201.2981},
 primaryClass = {astro-ph.HE},
       adsurl = {https://ui.adsabs.harvard.edu/abs/2012ApJS..199...19G}
}

@ARTICLE{Kuiper2011ATel,
       author = {{Kuiper}, L. and {Hermsen}, W.},
        title = "{The spin-down of Swift J1834.9-0846: confirmation as magnetar}",
      journal = {The Astronomer's Telegram},
         year = 2011,
        month = aug,
       volume = {3577},
        pages = {1},
       adsurl = {https://ui.adsabs.harvard.edu/abs/2011ATel.3577....1K}
}

@ARTICLE{Gogus2011ATel,
       author = {{Gogus}, Ersin and {Kouveliotou}, Chryssa},
        title = "{RXTE Discovery of the Spin Period of Swift J1834.9-0846}",
      journal = {The Astronomer's Telegram},
         year = 2011,
        month = aug,
       volume = {3542},
        pages = {1},
       adsurl = {https://ui.adsabs.harvard.edu/abs/2011ATel.3542....1G}
}

@ARTICLE{Kargaltsev2012ApJ,
       author = {{Kargaltsev}, Oleg and {Kouveliotou}, Chryssa and {Pavlov}, George G. and {G{\"o}{\v{g}}{\"u}{\c{s}}}, Ersin and {Lin}, Lin and {Wachter}, Stefanie and {Griffith}, Roger L. and {Kaneko}, Yuki and {Younes}, George},
        title = "{X-Ray Observations of the New Unusual Magnetar Swift J1834.9-0846}",
      journal = {\apj},
         year = 2012,
        month = mar,
       volume = {748},
       number = {1},
          eid = {26},
        pages = {26},
          doi = {10.1088/0004-637X/748/1/26},
archivePrefix = {arXiv},
       eprint = {1201.1894},
 primaryClass = {astro-ph.HE},
       adsurl = {https://ui.adsabs.harvard.edu/abs/2012ApJ...748...26K}
}

@ARTICLE{Torres2017ApJ,
       author = {{Torres}, Diego F.},
        title = "{Rotationally Powered Magnetar Nebula around Swift J1834.9-0846}",
      journal = {\apj},
         year = 2017,
        month = jan,
       volume = {835},
       number = {1},
          eid = {54},
        pages = {54},
          doi = {10.3847/1538-4357/835/1/54},
archivePrefix = {arXiv},
       eprint = {1612.02835},
 primaryClass = {astro-ph.HE},
       adsurl = {https://ui.adsabs.harvard.edu/abs/2017ApJ...835...54T}
}

@ARTICLE{Tian2007ApJ,
       author = {{Tian}, W.~W. and {Li}, Z. and {Leahy}, D.~A. and {Wang}, Q.~D.},
        title = "{VLA and XMM-Newton Observations of the SNR W41/TeV Gamma-Ray Source HESS J1834-087}",
      journal = {\apjl},
         year = 2007,
        month = mar,
       volume = {657},
       number = {1},
        pages = {L25-L28},
          doi = {10.1086/512544},
archivePrefix = {arXiv},
       eprint = {astro-ph/0612296},
 primaryClass = {astro-ph},
       adsurl = {https://ui.adsabs.harvard.edu/abs/2007ApJ...657L..25T}
}

@ARTICLE{Leahy2008AJ,
       author = {{Leahy}, D.~A. and {Tian}, W.~W.},
        title = "{The Distances of SNR W41 and Overlapping H II Regions}",
      journal = {\aj},
         year = 2008,
        month = jan,
       volume = {135},
       number = {1},
        pages = {167-172},
          doi = {10.1088/0004-6256/135/1/167},
archivePrefix = {arXiv},
       eprint = {0708.3377},
 primaryClass = {astro-ph},
       adsurl = {https://ui.adsabs.harvard.edu/abs/2008AJ....135..167L}
}

@ARTICLE{Espinoza2011MNRAS,
       author = {{Espinoza}, C.~M. and {Lyne}, A.~G. and {Stappers}, B.~W. and {Kramer}, M.},
        title = "{A study of 315 glitches in the rotation of 102 pulsars}",
      journal = {\mnras},
         year = 2011,
        month = jun,
       volume = {414},
       number = {2},
        pages = {1679-1704},
          doi = {10.1111/j.1365-2966.2011.18503.x},
archivePrefix = {arXiv},
       eprint = {1102.1743},
 primaryClass = {astro-ph.HE},
       adsurl = {https://ui.adsabs.harvard.edu/abs/2011MNRAS.414.1679E}
}

@ARTICLE{Lyne1993MNRAS,
       author = {{Lyne}, A.~G. and {Pritchard}, R.~S. and {Graham Smith}, F.},
        title = "{23 years of Crab pulsar rotational history.}",
      journal = {\mnras},
         year = 1993,
        month = dec,
       volume = {265},
        pages = {1003-1012},
          doi = {10.1093/mnras/265.4.1003},
       adsurl = {https://ui.adsabs.harvard.edu/abs/1993MNRAS.265.1003L}
}

@ARTICLE{Livingstone2007ApSS,
       author = {{Livingstone}, Margaret A. and {Kaspi}, Victoria M. and {Gavriil}, Fotis P. and {Manchester}, Richard N. and {Gotthelf}, E.~V.~G. and {Kuiper}, Lucien},
        title = "{New phase-coherent measurements of pulsar braking indices}",
      journal = {\apss},
         year = 2007,
        month = apr,
       volume = {308},
       number = {1-4},
        pages = {317-323},
          doi = {10.1007/s10509-007-9320-3},
archivePrefix = {arXiv},
       eprint = {astro-ph/0702196},
 primaryClass = {astro-ph},
       adsurl = {https://ui.adsabs.harvard.edu/abs/2007Ap&SS.308..317L}
}

@ARTICLE{Weltevred2011MNRAS,
       author = {{Weltevrede}, Patrick and {Johnston}, Simon and {Espinoza}, Crist{\'o}bal M.},
        title = "{The glitch-induced identity changes of PSR J1119-6127}",
      journal = {\mnras},
         year = 2011,
        month = mar,
       volume = {411},
       number = {3},
        pages = {1917-1934},
          doi = {10.1111/j.1365-2966.2010.17821.x},
archivePrefix = {arXiv},
       eprint = {1010.0857},
 primaryClass = {astro-ph.SR},
       adsurl = {https://ui.adsabs.harvard.edu/abs/2011MNRAS.411.1917W}
}

@ARTICLE{Lyne1996Natur,
       author = {{Lyne}, A.~G. and {Pritchard}, R.~S. and {Graham-Smith}, F. and {Camilo}, F.},
        title = "{Very low braking index for the Vela pulsar}",
      journal = {\nat},
         year = 1996,
        month = jun,
       volume = {381},
       number = {6582},
        pages = {497-498},
          doi = {10.1038/381497a0},
       adsurl = {https://ui.adsabs.harvard.edu/abs/1996Natur.381..497L}
}

@ARTICLE{Roy2012MNRAS,
       author = {{Roy}, Jayanta and {Gupta}, Yashwant and {Lewandowski}, Wojciech},
        title = "{Observations of four glitches in the young pulsar J1833-1034 and study of its glitch activity}",
      journal = {\mnras},
         year = 2012,
        month = aug,
       volume = {424},
       number = {3},
        pages = {2213-2221},
          doi = {10.1111/j.1365-2966.2012.21380.x},
archivePrefix = {arXiv},
       eprint = {1205.6264},
 primaryClass = {astro-ph.SR},
       adsurl = {https://ui.adsabs.harvard.edu/abs/2012MNRAS.424.2213R}
}

@ARTICLE{Araujo2016JCAP,
       author = {{de Araujo}, Jos{\'e} C.~N. and {Coelho}, Jaziel G. and {Costa}, Cesar A.},
        title = "{Gravitational wave emission by the high braking index pulsar PSR J1640-4631}",
      journal = {\jcap},
         year = 2016,
        month = jul,
       volume = {2016},
       number = {7},
          eid = {023},
        pages = {023},
          doi = {10.1088/1475-7516/2016/07/023},
archivePrefix = {arXiv},
       eprint = {1603.05975},
 primaryClass = {astro-ph.HE},
       adsurl = {https://ui.adsabs.harvard.edu/abs/2016JCAP...07..023D}
}

@ARTICLE{Kou2015MNRAS,
       author = {{Kou}, F.~F. and {Tong}, H.},
        title = "{Rotational evolution of the Crab pulsar in the wind braking model}",
      journal = {\mnras},
         year = 2015,
        month = jun,
       volume = {450},
       number = {2},
        pages = {1990-1998},
          doi = {10.1093/mnras/stv734},
archivePrefix = {arXiv},
       eprint = {1501.01534},
 primaryClass = {astro-ph.HE},
       adsurl = {https://ui.adsabs.harvard.edu/abs/2015MNRAS.450.1990K}
}

@ARTICLE{Xu2001ApJ,
       author = {{Xu}, R.~X. and {Qiao}, G.~J.},
        title = "{Pulsar Braking Index: A Test of Emission Models?}",
      journal = {\apjl},
         year = 2001,
        month = nov,
       volume = {561},
       number = {1},
        pages = {L85-L88},
          doi = {10.1086/324381},
archivePrefix = {arXiv},
       eprint = {astro-ph/0108235},
 primaryClass = {astro-ph},
       adsurl = {https://ui.adsabs.harvard.edu/abs/2001ApJ...561L..85X}
}

@ARTICLE{Gao2016MNRAS,
       author = {{Gao}, Z.~F. and {Li}, X. -D. and {Wang}, N. and {Yuan}, J.~P. and {Wang}, P. and {Peng}, Q.~H. and {Du}, Y.~J.},
        title = "{Constraining the braking indices of magnetars}",
      journal = {\mnras},
         year = 2016,
        month = feb,
       volume = {456},
       number = {1},
        pages = {55-65},
          doi = {10.1093/mnras/stv2465},
archivePrefix = {arXiv},
       eprint = {1505.07013},
 primaryClass = {astro-ph.HE},
       adsurl = {https://ui.adsabs.harvard.edu/abs/2016MNRAS.456...55G}
}

@ARTICLE{Younes2016ApJ,
       author = {{Younes}, G. and {Kouveliotou}, C. and {Kargaltsev}, O. and {Gill}, R. and {Granot}, J. and {Watts}, A.~L. and {Gelfand}, J. and {Baring}, M.~G. and {Harding}, A. and {Pavlov}, G.~G. and {van der Horst}, A.~J. and {Huppenkothen}, D. and {G{\"o}{\u{g}}{\"u}{\c{s}}}, E. and {Lin}, L. and {Roberts}, O.~J.},
        title = "{The Wind Nebula around Magnetar Swift J1834.9-0846}",
      journal = {\apj},
         year = 2016,
        month = jun,
       volume = {824},
       number = {2},
          eid = {138},
        pages = {138},
          doi = {10.3847/0004-637X/824/2/138},
archivePrefix = {arXiv},
       eprint = {1604.06472},
 primaryClass = {astro-ph.HE},
       adsurl = {https://ui.adsabs.harvard.edu/abs/2016ApJ...824..138Y}
}

@ARTICLE{Younes2012ApJ,
       author = {{Younes}, G. and {Kouveliotou}, C. and {Kargaltsev}, O. and {Pavlov}, G.~G. and {G{\"o}{\v{g}}{\"u}{\c{s}}}, E. and {Wachter}, S.},
        title = "{XMM-Newton View of Swift J1834.9-0846 and Its Magnetar Wind Nebula}",
      journal = {\apj},
         year = 2012,
        month = sep,
       volume = {757},
       number = {1},
          eid = {39},
        pages = {39},
          doi = {10.1088/0004-637X/757/1/39},
archivePrefix = {arXiv},
       eprint = {1206.3330},
 primaryClass = {astro-ph.HE},
       adsurl = {https://ui.adsabs.harvard.edu/abs/2012ApJ...757...39Y}
}

@ARTICLE{Jones1976ApSS,
       author = {{Jones}, P.~B.},
        title = "{The Secular Variation of Pulsar Magnetic Dipole Moments}",
      journal = {\apss},
         year = 1976,
        month = dec,
       volume = {45},
       number = {2},
        pages = {369-381},
          doi = {10.1007/BF00642671},
       adsurl = {https://ui.adsabs.harvard.edu/abs/1976Ap&SS..45..369J}
}

@ARTICLE{Cutler2002PhRvD,
       author = {{Cutler}, Curt},
        title = "{Gravitational waves from neutron stars with large toroidal B fields}",
      journal = {\prd},
         year = 2002,
        month = oct,
       volume = {66},
       number = {8},
          eid = {084025},
        pages = {084025},
          doi = {10.1103/PhysRevD.66.084025},
archivePrefix = {arXiv},
       eprint = {gr-qc/0206051},
 primaryClass = {gr-qc},
       adsurl = {https://ui.adsabs.harvard.edu/abs/2002PhRvD..66h4025C}
}

@ARTICLE{Alpar1988ApJ,
       author = {{Alpar}, M.~A. and {Sauls}, J.~A.},
        title = "{On the Dynamical Coupling between the Superfluid Interior and the Crust of a Neutron Star}",
      journal = {\apj},
         year = 1988,
        month = apr,
       volume = {327},
        pages = {723},
          doi = {10.1086/166228},
       adsurl = {https://ui.adsabs.harvard.edu/abs/1988ApJ...327..723A}
}

@ARTICLE{Cheng2019PhRvD,
       author = {{Cheng}, Quan and {Zhang}, Shuang-Nan and {Zheng}, Xiao-Ping and {Fan}, Xi-Long},
        title = "{What can PSR J1640-4631 tell us about the internal physics of this neutron star?}",
      journal = {\prd},
         year = 2019,
        month = apr,
       volume = {99},
       number = {8},
          eid = {083011},
        pages = {083011},
          doi = {10.1103/PhysRevD.99.083011},
archivePrefix = {arXiv},
       eprint = {1904.06570},
 primaryClass = {astro-ph.HE},
       adsurl = {https://ui.adsabs.harvard.edu/abs/2019PhRvD..99h3011C}
}

@ARTICLE{Hu2023RAA,
       author = {{Hu}, Fang-Yuan and {Cheng}, Quan and {Zheng}, Xiao-Ping and {Wang}, Jia-Qian and {Yan}, Yu-Long and {Wang}, Jia-Yu and {Luo}, Tian-Yu},
        title = "{Probing the Internal Physics of Neutron Stars through the Observed Braking Indices and Magnetic Tilt Angles of Several Young Pulsars}",
      journal = {Research in Astronomy and Astrophysics},
         year = 2023,
        month = may,
       volume = {23},
       number = {5},
          eid = {055020},
        pages = {055020},
          doi = {10.1088/1674-4527/accb7b},
       adsurl = {https://ui.adsabs.harvard.edu/abs/2023RAA....23e5020H}
}

@ARTICLE{Kaspi2017ARAA,
       author = {{Kaspi}, Victoria M. and {Beloborodov}, Andrei M.},
        title = "{Magnetars}",
      journal = {\araa},
         year = 2017,
        month = aug,
       volume = {55},
       number = {1},
        pages = {261-301},
          doi = {10.1146/annurev-astro-081915-023329},
archivePrefix = {arXiv},
       eprint = {1703.00068},
 primaryClass = {astro-ph.HE},
       adsurl = {https://ui.adsabs.harvard.edu/abs/2017ARA&A..55..261K}
}

@ARTICLE{Bochenek2020Natur,
       author = {{Bochenek}, C.~D. and {Ravi}, V. and {Belov}, K.~V. and {Hallinan}, G. and {Kocz}, J. and {Kulkarni}, S.~R. and {McKenna}, D.~L.},
        title = "{A fast radio burst associated with a Galactic magnetar}",
      journal = {\nat},
         year = 2020,
        month = nov,
       volume = {587},
       number = {7832},
        pages = {59-62},
          doi = {10.1038/s41586-020-2872-x},
archivePrefix = {arXiv},
       eprint = {2005.10828},
 primaryClass = {astro-ph.HE},
       adsurl = {https://ui.adsabs.harvard.edu/abs/2020Natur.587...59B}
}

@ARTICLE{Turolla2015RPPh,
       author = {{Turolla}, R. and {Zane}, S. and {Watts}, A.~L.},
        title = "{Magnetars: the physics behind observations. A review}",
      journal = {Reports on Progress in Physics},
         year = 2015,
        month = nov,
       volume = {78},
       number = {11},
          eid = {116901},
        pages = {116901},
          doi = {10.1088/0034-4885/78/11/116901},
archivePrefix = {arXiv},
       eprint = {1507.02924},
 primaryClass = {astro-ph.HE},
       adsurl = {https://ui.adsabs.harvard.edu/abs/2015RPPh...78k6901T}
}

@ARTICLE{Mereghetti2015SSRv,
       author = {{Mereghetti}, Sandro and {Pons}, Jos{\'e} A. and {Melatos}, Andrew},
        title = "{Magnetars: Properties, Origin and Evolution}",
      journal = {\ssr},
         year = 2015,
        month = oct,
       volume = {191},
       number = {1-4},
        pages = {315-338},
          doi = {10.1007/s11214-015-0146-y},
archivePrefix = {arXiv},
       eprint = {1503.06313},
 primaryClass = {astro-ph.HE},
       adsurl = {https://ui.adsabs.harvard.edu/abs/2015SSRv..191..315M}
}

@ARTICLE{Thompson2004ApJ,
       author = {{Thompson}, Todd A. and {Chang}, Philip and {Quataert}, Eliot},
        title = "{Magnetar Spin-Down, Hyperenergetic Supernovae, and Gamma-Ray Bursts}",
      journal = {\apj},
         year = 2004,
        month = aug,
       volume = {611},
       number = {1},
        pages = {380-393},
          doi = {10.1086/421969},
archivePrefix = {arXiv},
       eprint = {astro-ph/0401555},
 primaryClass = {astro-ph},
       adsurl = {https://ui.adsabs.harvard.edu/abs/2004ApJ...611..380T}
}

@ARTICLE{Ferdman2015ApJ,
       author = {{Ferdman}, R.~D. and {Archibald}, R.~F. and {Kaspi}, V.~M.},
        title = "{Long-term Timing and Emission Behavior of the Young Crab-like Pulsar PSR B0540-69}",
      journal = {\apj},
         year = 2015,
        month = oct,
       volume = {812},
       number = {2},
          eid = {95},
        pages = {95},
          doi = {10.1088/0004-637X/812/2/95},
archivePrefix = {arXiv},
       eprint = {1506.00182},
 primaryClass = {astro-ph.SR},
       adsurl = {https://ui.adsabs.harvard.edu/abs/2015ApJ...812...95F}
}

@ARTICLE{Ostriker1969ApJ,
       author = {{Ostriker}, J.~P. and {Gunn}, J.~E.},
        title = "{On the Nature of Pulsars. I. Theory}",
      journal = {\apj},
         year = 1969,
        month = sep,
       volume = {157},
        pages = {1395},
          doi = {10.1086/150160},
       adsurl = {https://ui.adsabs.harvard.edu/abs/1969ApJ...157.1395O}
}

@ARTICLE{Eksi2016ApJ,
       author = {{Ek{\c{s}}i}, K.~Y. and {Anda{\c{c}}}, I.~C. and {{\c{C}}{\i}k{\i}nto{\u{g}}lu}, S. and {G{\"u}gercino{\u{g}}lu}, E. and {Vahdat Motlagh}, A. and {K{\i}z{\i}ltan}, B.},
        title = "{The Inclination Angle and Evolution of the Braking Index of Pulsars with Plasma-filled Magnetosphere: Application to the High Braking Index of PSR J1640-4631}",
      journal = {\apj},
         year = 2016,
        month = may,
       volume = {823},
       number = {1},
          eid = {34},
        pages = {34},
          doi = {10.3847/0004-637X/823/1/34},
archivePrefix = {arXiv},
       eprint = {1603.01487},
 primaryClass = {astro-ph.HE},
       adsurl = {https://ui.adsabs.harvard.edu/abs/2016ApJ...823...34E}
}

@ARTICLE{Gao2017ApJ,
       author = {{Gao}, Zhi-Fu and {Wang}, Na and {Shan}, Hao and {Li}, Xiang-Dong and {Wang}, Wei},
        title = "{The Dipole Magnetic Field and Spin-down Evolutions of the High Braking Index Pulsar PSR J1640-4631}",
      journal = {\apj},
         year = 2017,
        month = nov,
       volume = {849},
       number = {1},
          eid = {19},
        pages = {19},
          doi = {10.3847/1538-4357/aa8f49},
archivePrefix = {arXiv},
       eprint = {1709.03459},
 primaryClass = {astro-ph.HE},
       adsurl = {https://ui.adsabs.harvard.edu/abs/2017ApJ...849...19G}
}

@ARTICLE{Cutler2000PhRvD,
       author = {{Cutler}, Curt and {Jones}, David Ian},
        title = "{Gravitational wave damping of neutron star wobble}",
      journal = {\prd},
         year = 2000,
        month = dec,
       volume = {63},
       number = {2},
          eid = {024002},
        pages = {024002},
          doi = {10.1103/PhysRevD.63.024002},
archivePrefix = {arXiv},
       eprint = {gr-qc/0008021},
 primaryClass = {gr-qc},
       adsurl = {https://ui.adsabs.harvard.edu/abs/2000PhRvD..63b4002C}
}

@ARTICLE{Philippov2014MNRAS,
       author = {{Philippov}, Alexander and {Tchekhovskoy}, Alexander and {Li}, Jason G.},
        title = "{Time evolution of pulsar obliquity angle from 3D simulations of magnetospheres}",
      journal = {\mnras},
         year = 2014,
        month = jul,
       volume = {441},
       number = {3},
        pages = {1879-1887},
          doi = {10.1093/mnras/stu591},
archivePrefix = {arXiv},
       eprint = {1311.1513},
 primaryClass = {astro-ph.HE},
       adsurl = {https://ui.adsabs.harvard.edu/abs/2014MNRAS.441.1879P}
}

@ARTICLE{Spitkovsky2006ApJ,
       author = {{Spitkovsky}, Anatoly},
        title = "{Time-dependent Force-free Pulsar Magnetospheres: Axisymmetric and Oblique Rotators}",
      journal = {\apjl},
         year = 2006,
        month = sep,
       volume = {648},
       number = {1},
        pages = {L51-L54},
          doi = {10.1086/507518},
archivePrefix = {arXiv},
       eprint = {astro-ph/0603147},
 primaryClass = {astro-ph},
       adsurl = {https://ui.adsabs.harvard.edu/abs/2006ApJ...648L..51S}
}

@ARTICLE{Riles2023LRR,
       author = {{Riles}, Keith},
        title = "{Searches for continuous-wave gravitational radiation}",
      journal = {Living Reviews in Relativity},
         year = 2023,
        month = dec,
       volume = {26},
       number = {1},
          eid = {3},
        pages = {3},
          doi = {10.1007/s41114-023-00044-3},
archivePrefix = {arXiv},
       eprint = {2206.06447},
 primaryClass = {astro-ph.HE},
       adsurl = {https://ui.adsabs.harvard.edu/abs/2023LRR....26....3R}
}

@ARTICLE{Bonazzola1996AA,
       author = {{Bonazzola}, S. and {Gourgoulhon}, E.},
        title = "{Gravitational waves from pulsars: emission by the magnetic-field-induced distortion.}",
      journal = {\aap},
         year = 1996,
        month = aug,
       volume = {312},
        pages = {675-690},
          doi = {10.48550/arXiv.astro-ph/9602107},
archivePrefix = {arXiv},
       eprint = {astro-ph/9602107},
 primaryClass = {astro-ph},
       adsurl = {https://ui.adsabs.harvard.edu/abs/1996A&A...312..675B}
}

@ARTICLE{Haskell2008MNRAS,
       author = {{Haskell}, B. and {Samuelsson}, L. and {Glampedakis}, K. and {Andersson}, N.},
        title = "{Modelling magnetically deformed neutron stars}",
      journal = {\mnras},
         year = 2008,
        month = mar,
       volume = {385},
       number = {1},
        pages = {531-542},
          doi = {10.1111/j.1365-2966.2008.12861.x},
archivePrefix = {arXiv},
       eprint = {0705.1780},
 primaryClass = {astro-ph},
       adsurl = {https://ui.adsabs.harvard.edu/abs/2008MNRAS.385..531H}
}

@ARTICLE{Tong2017ApJ,
       author = {{Tong}, H. and {Kou}, F.~F.},
        title = "{Possible Evolution of the Pulsar Braking Index from Larger than Three to About One}",
      journal = {\apj},
         year = 2017,
        month = mar,
       volume = {837},
       number = {2},
          eid = {117},
        pages = {117},
          doi = {10.3847/1538-4357/aa60c6},
archivePrefix = {arXiv},
       eprint = {1604.01231},
 primaryClass = {astro-ph.HE},
       adsurl = {https://ui.adsabs.harvard.edu/abs/2017ApJ...837..117T}
}

@ARTICLE{Ou2016MNRAS,
       author = {{Ou}, Z.~W. and {Tong}, H. and {Kou}, F.~F. and {Ding}, G.~Q.},
        title = "{Fluctuating neutron star magnetosphere: braking indices of eight pulsars, frequency second derivatives of 222 pulsars and 15 magnetars}",
      journal = {\mnras},
         year = 2016,
        month = apr,
       volume = {457},
       number = {4},
        pages = {3922-3933},
          doi = {10.1093/mnras/stw227},
archivePrefix = {arXiv},
       eprint = {1512.01679},
 primaryClass = {astro-ph.HE},
       adsurl = {https://ui.adsabs.harvard.edu/abs/2016MNRAS.457.3922O}
}

@ARTICLE{Yue2007AdSpR,
       author = {{Yue}, Y.~L. and {Xu}, R.~X. and {Zhu}, W.~W.},
        title = "{What can the braking indices tell us about the nature of pulsars?}",
      journal = {Advances in Space Research},
         year = 2007,
        month = jan,
       volume = {40},
       number = {10},
        pages = {1491-1497},
          doi = {10.1016/j.asr.2007.08.016},
archivePrefix = {arXiv},
       eprint = {astro-ph/0611022},
 primaryClass = {astro-ph},
       adsurl = {https://ui.adsabs.harvard.edu/abs/2007AdSpR..40.1491Y}
}

@ARTICLE{Ruderman1975ApJ,
       author = {{Ruderman}, M.~A. and {Sutherland}, P.~G.},
        title = "{Theory of pulsars: polar gaps, sparks, and coherent microwave radiation.}",
      journal = {\apj},
         year = 1975,
        month = feb,
       volume = {196},
        pages = {51-72},
          doi = {10.1086/153393},
       adsurl = {https://ui.adsabs.harvard.edu/abs/1975ApJ...196...51R}
}

@ARTICLE{Camilo2007ApJ,
       author = {{Camilo}, F. and {Reynolds}, J. and {Johnston}, S. and {Halpern}, J.~P. and {Ransom}, S.~M. and {van Straten}, W.},
        title = "{Polarized Radio Emission from the Magnetar XTE J1810-197}",
      journal = {\apjl},
         year = 2007,
        month = apr,
       volume = {659},
       number = {1},
        pages = {L37-L40},
          doi = {10.1086/516630},
archivePrefix = {arXiv},
       eprint = {astro-ph/0702616},
 primaryClass = {astro-ph},
       adsurl = {https://ui.adsabs.harvard.edu/abs/2007ApJ...659L..37C}
}

@ARTICLE{Camilo2008ApJ,
       author = {{Camilo}, F. and {Reynolds}, J. and {Johnston}, S. and {Halpern}, J.~P. and {Ransom}, S.~M.},
        title = "{The Magnetar 1E 1547.0-5408: Radio Spectrum, Polarimetry, and Timing}",
      journal = {\apj},
         year = 2008,
        month = may,
       volume = {679},
       number = {1},
        pages = {681-686},
          doi = {10.1086/587054},
archivePrefix = {arXiv},
       eprint = {0802.0494},
 primaryClass = {astro-ph},
       adsurl = {https://ui.adsabs.harvard.edu/abs/2008ApJ...679..681C}
}

@ARTICLE{Lower2021MNRAS,
       author = {{Lower}, M.~E. and {Johnston}, S. and {Shannon}, R.~M. and {Bailes}, M. and {Camilo}, F.},
        title = "{The dynamic magnetosphere of Swift J1818.0-1607}",
      journal = {\mnras},
         year = 2021,
        month = mar,
       volume = {502},
       number = {1},
        pages = {127-139},
          doi = {10.1093/mnras/staa3789},
archivePrefix = {arXiv},
       eprint = {2011.12463},
 primaryClass = {astro-ph.HE},
       adsurl = {https://ui.adsabs.harvard.edu/abs/2021MNRAS.502..127L}
}

@ARTICLE{Levin2012MNRAS,
       author = {{Levin}, L. and {Bailes}, M. and {Bates}, S.~D. and {Bhat}, N.~D.~R. and {Burgay}, M. and {Burke-Spolaor}, S. and {D'Amico}, N. and {Johnston}, S. and {Keith}, M.~J. and {Kramer}, M. and {Milia}, S. and {Possenti}, A. and {Stappers}, B. and {van Straten}, W.},
        title = "{Radio emission evolution, polarimetry and multifrequency single pulse analysis of the radio magnetar PSR J1622-4950}",
      journal = {\mnras},
         year = 2012,
        month = may,
       volume = {422},
       number = {3},
        pages = {2489-2500},
          doi = {10.1111/j.1365-2966.2012.20807.x},
archivePrefix = {arXiv},
       eprint = {1204.2045},
 primaryClass = {astro-ph.HE},
       adsurl = {https://ui.adsabs.harvard.edu/abs/2012MNRAS.422.2489L}
}

@ARTICLE{Peng2024ApJ,
       author = {{Peng}, Han-Long and {Ge}, Ming-Yu and {Weng}, Shan-Shan and {Zhao}, Qing-Chang and {Ye}, Wen-Tao and {Zhang}, Liang and {Qi}, Li-Qiang and {Tuo}, You-Li},
        title = "{Polarized X-Rays Detected from the Anomalous X-Ray Pulsar 1E 2259+586}",
      journal = {\apj},
         year = 2024,
        month = jan,
       volume = {961},
       number = {1},
          eid = {106},
        pages = {106},
          doi = {10.3847/1538-4357/ad1512},
       adsurl = {https://ui.adsabs.harvard.edu/abs/2024ApJ...961..106P}
}

@ARTICLE{Heyl2024MNRAS,
       author = {{Heyl}, Jeremy and {Taverna}, Roberto and {Turolla}, Roberto and {Israel}, Gian Luca and {Ng}, Mason and {K{\i}rm{\i}z{\i}bayrak}, Demet and {Gonz{\'a}lez-Caniulef}, Denis and {Caiazzo}, Ilaria and {Zane}, Silvia and {Ehlert}, Steven R. and {Negro}, Michela and {Agudo}, Iv{\'a}n and {Antonelli}, Lucio Angelo and {Bachetti}, Matteo and {Baldini}, Luca and {Baumgartner}, Wayne H. and {Bellazzini}, Ronaldo and {Bianchi}, Stefano and {Bongiorno}, Stephen D. and {Bonino}, Raffaella and {Brez}, Alessandro and {Bucciantini}, Niccol{\`o} and {Capitanio}, Fiamma and {Castellano}, Simone and {Cavazzuti}, Elisabetta and {Chen}, Chien-Ting and {Ciprini}, Stefano and {Costa}, Enrico and {De Rosa}, Alessandra and {Del Monte}, Ettore and {Di Gesu}, Laura and {Di Lalla}, Niccol{\`o} and {Di Marco}, Alessandro and {Donnarumma}, Immacolata and {Doroshenko}, Victor and {Dov{\v{c}}iak}, Michal and {Enoto}, Teruaki and {Evangelista}, Yuri and {Fabiani}, Sergio and {Ferrazzoli}, Riccardo and {Garcia}, Javier A. and {Gunji}, Shuichi and {Hayashida}, Kiyoshi and {Iwakiri}, Wataru and {Jorstad}, Svetlana G. and {Kaaret}, Philip and {Karas}, Vladimir and {Kislat}, Fabian and {Kitaguchi}, Takao and {Kolodziejczak}, Jeffery J. and {Krawczynski}, Henric and {Monaca}, Fabio La and {Latronico}, Luca and {Liodakis}, Ioannis and {Maldera}, Simone and {Manfreda}, Alberto and {Marin}, Fr{\'e}d{\'e}ric and {Marinucci}, Andrea and {Marscher}, Alan P. and {Marshall}, Herman L. and {Massaro}, Francesco and {Matt}, Giorgio and {Mitsuishi}, Ikuyuki and {Mizuno}, Tsunefumi and {Muleri}, Fabio and {Ng}, C. -Y. and {O'Dell}, Stephen L. and {Omodei}, Nicola and {Oppedisano}, Chiara and {Papitto}, Alessandro and {Pavlov}, George G. and {Peirson}, Abel Lawrence and {Perri}, Matteo and {Pesce-Rollins}, Melissa and {Petrucci}, Pierre-Olivier and {Pilia}, Maura and {Possenti}, Andrea and {Poutanen}, Juri and {Puccetti}, Simonetta and {Ramsey}, Brian D. and {Rankin}, John and {Ratheesh}, Ajay and {Roberts}, Oliver J. and {Romani}, Roger W. and {Sgr{\`o}}, Carmelo and {Slane}, Patrick and {Soffitta}, Paolo and {Spandre}, Gloria and {Swartz}, Douglas A. and {Tamagawa}, Toru and {Tavecchio}, Fabrizio and {Tawara}, Yuzuru and {Tennant}, Allyn F. and {Thomas}, Nicholas E. and {Tombesi}, Francesco and {Trois}, Alessio and {Tsygankov}, Sergey S. and {Vink}, Jacco and {Weisskopf}, Martin C. and {Wu}, Kinwah and {Xie}, Fei},
        title = "{The detection of polarized X-ray emission from the magnetar 1E 2259+586}",
      journal = {\mnras},
         year = 2024,
        month = feb,
       volume = {527},
       number = {4},
        pages = {12219-12231},
          doi = {10.1093/mnras/stad3680},
archivePrefix = {arXiv},
       eprint = {2311.03637},
 primaryClass = {astro-ph.HE},
       adsurl = {https://ui.adsabs.harvard.edu/abs/2024MNRAS.52712219H}
}

@ARTICLE{Zane2023ApJ,
       author = {{Zane}, Silvia and {Taverna}, Roberto and {Gonz{\'a}lez-Caniulef}, Denis and {Muleri}, Fabio and {Turolla}, Roberto and {Heyl}, Jeremy and {Uchiyama}, Keisuke and {Ng}, Mason and {Tamagawa}, Toru and {Caiazzo}, Ilaria and {Di Lalla}, Niccol{\`o} and {Marshall}, Herman L. and {Bachetti}, Matteo and {La Monaca}, Fabio and {Gau}, Ephraim and {Di Marco}, Alessandro and {Baldini}, Luca and {Negro}, Michela and {Omodei}, Nicola and {Rankin}, John and {Matt}, Giorgio and {Pavlov}, George G. and {Kitaguchi}, Takao and {Krawczynski}, Henric and {Kislat}, Fabian and {Kelly}, Ruth and {Agudo}, Iv{\'a}n and {Antonelli}, Lucio A. and {Baumgartner}, Wayne H. and {Bellazzini}, Ronaldo and {Bianchi}, Stefano and {Bongiorno}, Stephen D. and {Bonino}, Raffaella and {Brez}, Alessandro and {Bucciantini}, Niccol{\`o} and {Capitanio}, Fiamma and {Castellano}, Simone and {Cavazzuti}, Elisabetta and {Chen}, Chieng-Ting and {Ciprini}, Stefano and {Costa}, Enrico and {De Rosa}, Alessandra and {Del Monte}, Ettore and {Di Gesu}, Laura and {Donnarumma}, Immacolata and {Doroshenko}, Victor and {Dov{\v{c}}iak}, Michal and {Ehlert}, Steven R. and {Enoto}, Teruaki and {Evangelista}, Yuri and {Fabiani}, Sergio and {Ferrazzoli}, Riccardo and {Garcia}, Javier A. and {Gunji}, Shuichi and {Hayashida}, Kiyoshi and {Iwakiri}, Wataru and {Jorstad}, Svetlana G. and {Kaaret}, Philip and {Karas}, Vladimir and {Kolodziejczak}, Jeffery J. and {Latronico}, Luca and {Liodakis}, Ioannis and {Maldera}, Simone and {Manfreda}, Alberto and {Marin}, Fr{\'e}d{\'e}ric and {Marinucci}, Andrea and {Marscher}, Alan P. and {Massaro}, Francesco and {Mitsuishi}, Ikuyuki and {Mizuno}, Tsunefumi and {Ng}, C. -Y. and {O'Dell}, Stephen L. and {Oppedisano}, Chiara and {Papitto}, Alessandro and {Peirson}, Abel L. and {Perri}, Matteo and {Pesce-Rollins}, Melissa and {Petrucci}, Pierre-Olivier and {Pilia}, Maura and {Possenti}, Andrea and {Poutanen}, Juri and {Puccetti}, Simonetta and {Ramsey}, Brian D. and {Ratheesh}, Ajay and {Roberts}, Oliver J. and {Romani}, Roger W. and {Sgr{\'o}}, Carmelo and {Slane}, Patrick and {Soffitta}, Paolo and {Spandre}, Gloria and {Swartz}, Douglas A. and {Tavecchio}, Fabrizio and {Tawara}, Yuzuru and {Tennant}, Allyn F. and {Thomas}, Nicholas E. and {Tombesi}, Francesco and {Trois}, Alessio and {Tsygankov}, Sergey S. and {Vink}, Jacco and {Weisskopf}, Martin C. and {Wu}, Kinwah and {Xie}, Fei},
        title = "{A Strong X-Ray Polarization Signal from the Magnetar 1RXS J170849.0-400910}",
      journal = {\apjl},
         year = 2023,
        month = feb,
       volume = {944},
       number = {2},
          eid = {L27},
        pages = {L27},
          doi = {10.3847/2041-8213/acb703},
archivePrefix = {arXiv},
       eprint = {2301.12919},
 primaryClass = {astro-ph.HE},
       adsurl = {https://ui.adsabs.harvard.edu/abs/2023ApJ...944L..27Z}
}

@ARTICLE{Harding2017ApJ,
       author = {{Harding}, Alice K. and {Kalapotharakos}, Constantinos},
        title = "{Multiwavelength Polarization of Rotation-powered Pulsars}",
      journal = {\apj},
         year = 2017,
        month = may,
       volume = {840},
       number = {2},
          eid = {73},
        pages = {73},
          doi = {10.3847/1538-4357/aa6ead},
archivePrefix = {arXiv},
       eprint = {1704.06183},
 primaryClass = {astro-ph.HE},
       adsurl = {https://ui.adsabs.harvard.edu/abs/2017ApJ...840...73H}
}

@ARTICLE{Thompson1995MNRAS,
       author = {{Thompson}, Christopher and {Duncan}, Robert C.},
        title = "{The soft gamma repeaters as very strongly magnetized neutron stars - I. Radiative mechanism for outbursts}",
      journal = {\mnras},
         year = 1995,
        month = jul,
       volume = {275},
       number = {2},
        pages = {255-300},
          doi = {10.1093/mnras/275.2.255},
       adsurl = {https://ui.adsabs.harvard.edu/abs/1995MNRAS.275..255T}
}

@ARTICLE{Philippov2022ARAA,
       author = {{Philippov}, A. and {Kramer}, M.},
        title = "{Pulsar Magnetospheres and Their Radiation}",
      journal = {\araa},
         year = 2022,
        month = aug,
       volume = {60},
        pages = {495-558},
          doi = {10.1146/annurev-astro-052920-112338},
       adsurl = {https://ui.adsabs.harvard.edu/abs/2022ARA&A..60..495P}
}

@ARTICLE{Cerutti2015MNRAS,
       author = {{Cerutti}, Beno{\^\i}t and {Philippov}, Alexander and {Parfrey}, Kyle and {Spitkovsky}, Anatoly},
        title = "{Particle acceleration in axisymmetric pulsar current sheets}",
      journal = {\mnras},
         year = 2015,
        month = mar,
       volume = {448},
       number = {1},
        pages = {606-619},
          doi = {10.1093/mnras/stv042},
archivePrefix = {arXiv},
       eprint = {1410.3757},
 primaryClass = {astro-ph.HE},
       adsurl = {https://ui.adsabs.harvard.edu/abs/2015MNRAS.448..606C}
}

@ARTICLE{Vigano2013MNRAS,
       author = {{Vigan{\`o}}, D. and {Rea}, N. and {Pons}, J.~A. and {Perna}, R. and {Aguilera}, D.~N. and {Miralles}, J.~A.},
        title = "{Unifying the observational diversity of isolated neutron stars via magneto-thermal evolution models}",
      journal = {\mnras},
         year = 2013,
        month = sep,
       volume = {434},
       number = {1},
        pages = {123-141},
          doi = {10.1093/mnras/stt1008},
archivePrefix = {arXiv},
       eprint = {1306.2156},
 primaryClass = {astro-ph.SR},
       adsurl = {https://ui.adsabs.harvard.edu/abs/2013MNRAS.434..123V}
}

@ARTICLE{Harding2002ApJ,
       author = {{Harding}, Alice K. and {Muslimov}, Alexander G.},
        title = "{Pulsar Polar Cap Heating and Surface Thermal X-Ray Emission. II. Inverse Compton Radiation Pair Fronts}",
      journal = {\apj},
         year = 2002,
        month = apr,
       volume = {568},
       number = {2},
        pages = {862-877},
          doi = {10.1086/338985},
archivePrefix = {arXiv},
       eprint = {astro-ph/0112392},
 primaryClass = {astro-ph},
       adsurl = {https://ui.adsabs.harvard.edu/abs/2002ApJ...568..862H}
}

@ARTICLE{Medin2007MNRAS,
       author = {{Medin}, Zach and {Lai}, Dong},
        title = "{Condensed surfaces of magnetic neutron stars, thermal surface emission, and particle acceleration above pulsar polar caps}",
      journal = {\mnras},
         year = 2007,
        month = dec,
       volume = {382},
       number = {4},
        pages = {1833-1852},
          doi = {10.1111/j.1365-2966.2007.12492.x},
       adsurl = {https://ui.adsabs.harvard.edu/abs/2007MNRAS.382.1833M}
}

@ARTICLE{Kalapotharakos2018ApJ,
       author = {{Kalapotharakos}, Constantinos and {Brambilla}, Gabriele and {Timokhin}, Andrey and {Harding}, Alice K. and {Kazanas}, Demosthenes},
        title = "{Three-dimensional Kinetic Pulsar Magnetosphere Models: Connecting to Gamma-Ray Observations}",
      journal = {\apj},
         year = 2018,
        month = apr,
       volume = {857},
       number = {1},
          eid = {44},
        pages = {44},
          doi = {10.3847/1538-4357/aab550},
archivePrefix = {arXiv},
       eprint = {1710.03170},
 primaryClass = {astro-ph.HE},
       adsurl = {https://ui.adsabs.harvard.edu/abs/2018ApJ...857...44K}
}

@ARTICLE{Younes2022ApJ,
       author = {{Younes}, George and {Lander}, Samuel K. and {Baring}, Matthew G. and {Enoto}, Teruaki and {Kouveliotou}, Chryssa and {Wadiasingh}, Zorawar and {Ho}, Wynn C.~G. and {Harding}, Alice K. and {Arzoumanian}, Zaven and {Gendreau}, Keith and {G{\"u}ver}, Tolga and {Hu}, Chin-Ping and {Malacaria}, Christian and {Ray}, Paul S. and {Strohmayer}, Tod E.},
        title = "{Pulse Peak Migration during the Outburst Decay of the Magnetar SGR 1830-0645: Crustal Motion and Magnetospheric Untwisting}",
      journal = {\apjl},
         year = 2022,
        month = jan,
       volume = {924},
       number = {2},
          eid = {L27},
        pages = {L27},
          doi = {10.3847/2041-8213/ac4700},
archivePrefix = {arXiv},
       eprint = {2201.05517},
 primaryClass = {astro-ph.HE},
       adsurl = {https://ui.adsabs.harvard.edu/abs/2022ApJ...924L..27Y}
}

@ARTICLE{Lander2013PhRvL,
       author = {{Lander}, S.~K.},
        title = "{Magnetic Fields in Superconducting Neutron Stars}",
      journal = {\prl},
         year = 2013,
        month = feb,
       volume = {110},
       number = {7},
          eid = {071101},
        pages = {071101},
          doi = {10.1103/PhysRevLett.110.071101},
archivePrefix = {arXiv},
       eprint = {1211.3912},
 primaryClass = {astro-ph.SR},
       adsurl = {https://ui.adsabs.harvard.edu/abs/2013PhRvL.110g1101L}
}

@ARTICLE{Akgun2008MNRAS,
       author = {{Akg{\"u}n}, T. and {Wasserman}, I.},
        title = "{Toroidal magnetic fields in type II superconducting neutron stars}",
      journal = {\mnras},
         year = 2008,
        month = feb,
       volume = {383},
       number = {4},
        pages = {1551-1580},
          doi = {10.1111/j.1365-2966.2007.12660.x},
archivePrefix = {arXiv},
       eprint = {0705.2195},
 primaryClass = {astro-ph},
       adsurl = {https://ui.adsabs.harvard.edu/abs/2008MNRAS.383.1551A}
}

@ARTICLE{Glampedakis2010MNRAS,
       author = {{Glampedakis}, K. and {Jones}, D.~I.},
        title = "{Implications of magnetar non-precession}",
      journal = {\mnras},
         year = 2010,
        month = jun,
       volume = {405},
       number = {1},
        pages = {L6-L10},
          doi = {10.1111/j.1745-3933.2010.00846.x},
archivePrefix = {arXiv},
       eprint = {1003.1208},
 primaryClass = {astro-ph.SR},
       adsurl = {https://ui.adsabs.harvard.edu/abs/2010MNRAS.405L...6G}
}

@ARTICLE{Pons2007PhRvL,
       author = {{Pons}, Jos{\'e} A. and {Link}, Bennett and {Miralles}, Juan A. and {Geppert}, Ulrich},
        title = "{Evidence for Heating of Neutron Stars by Magnetic-Field Decay}",
      journal = {\prl},
         year = 2007,
        month = feb,
       volume = {98},
       number = {7},
          eid = {071101},
        pages = {071101},
          doi = {10.1103/PhysRevLett.98.071101},
archivePrefix = {arXiv},
       eprint = {astro-ph/0607583},
 primaryClass = {astro-ph},
       adsurl = {https://ui.adsabs.harvard.edu/abs/2007PhRvL..98g1101P}
}

@ARTICLE{Goldreich1992ApJ,
       author = {{Goldreich}, Peter and {Reisenegger}, Andreas},
        title = "{Magnetic Field Decay in Isolated Neutron Stars}",
      journal = {\apj},
         year = 1992,
        month = aug,
       volume = {395},
        pages = {250},
          doi = {10.1086/171646},
       adsurl = {https://ui.adsabs.harvard.edu/abs/1992ApJ...395..250G}
}

@ARTICLE{Bransgrove2018MNRAS,
       author = {{Bransgrove}, Ashley and {Levin}, Yuri and {Beloborodov}, Andrei},
        title = "{Magnetic field evolution of neutron stars - I. Basic formalism, numerical techniques and first results}",
      journal = {\mnras},
         year = 2018,
        month = jan,
       volume = {473},
       number = {2},
        pages = {2771-2790},
          doi = {10.1093/mnras/stx2508},
archivePrefix = {arXiv},
       eprint = {1709.09167},
 primaryClass = {astro-ph.HE},
       adsurl = {https://ui.adsabs.harvard.edu/abs/2018MNRAS.473.2771B}
}

@ARTICLE{Cumming2004ApJ,
       author = {{Cumming}, Andrew and {Arras}, Phil and {Zweibel}, Ellen},
        title = "{Magnetic Field Evolution in Neutron Star Crusts Due to the Hall Effect and Ohmic Decay}",
      journal = {\apj},
         year = 2004,
        month = jul,
       volume = {609},
       number = {2},
        pages = {999-1017},
          doi = {10.1086/421324},
archivePrefix = {arXiv},
       eprint = {astro-ph/0402392},
 primaryClass = {astro-ph},
       adsurl = {https://ui.adsabs.harvard.edu/abs/2004ApJ...609..999C}
}

@software{Foreman2013ascl,
       author = {{Foreman-Mackey}, Daniel and {Conley}, Alex and {Meierjurgen Farr}, Will and {Hogg}, David W. and {Lang}, Dustin and {Marshall}, Phil and {Price-Whelan}, Adrian and {Sanders}, Jeremy and {Zuntz}, Joe},
        title = "{emcee: The MCMC Hammer}",
 howpublished = {Astrophysics Source Code Library, record ascl:1303.002},
         year = 2013,
        month = mar,
          eid = {ascl:1303.002},
       adsurl = {https://ui.adsabs.harvard.edu/abs/2013ascl.soft03002F}
}

@ARTICLE{Manchester2005AJ,
       author = {{Manchester}, R.~N. and {Hobbs}, G.~B. and {Teoh}, A. and {Hobbs}, M.},
        title = "{The Australia Telescope National Facility Pulsar Catalogue}",
      journal = {\aj},
         year = 2005,
        month = apr,
       volume = {129},
       number = {4},
        pages = {1993-2006},
          doi = {10.1086/428488},
archivePrefix = {arXiv},
       eprint = {astro-ph/0412641},
 primaryClass = {astro-ph},
       adsurl = {https://ui.adsabs.harvard.edu/abs/2005AJ....129.1993M}
}

@ARTICLE{Wette2023APh,
       author = {{Wette}, Karl},
        title = "{Searches for continuous gravitational waves from neutron stars: A twenty-year retrospective}",
      journal = {Astroparticle Physics},
         year = 2023,
        month = nov,
       volume = {153},
          eid = {102880},
        pages = {102880},
          doi = {10.1016/j.astropartphys.2023.102880},
archivePrefix = {arXiv},
       eprint = {2305.07106},
 primaryClass = {gr-qc},
       adsurl = {https://ui.adsabs.harvard.edu/abs/2023APh...15302880W}
}

@ARTICLE{Cheng2017PhRvD,
       author = {{Cheng}, Quan and {Zhang}, Shuang-Nan and {Zheng}, Xiao-Ping},
        title = "{Stochastic gravitational wave background from newly born massive magnetars: The role of a dense matter equation of state}",
      journal = {\prd},
         year = 2017,
        month = apr,
       volume = {95},
       number = {8},
          eid = {083003},
        pages = {083003},
          doi = {10.1103/PhysRevD.95.083003},
archivePrefix = {arXiv},
       eprint = {1704.02013},
 primaryClass = {astro-ph.HE},
       adsurl = {https://ui.adsabs.harvard.edu/abs/2017PhRvD..95h3003C}
}

@ARTICLE{Cheng2015MNRAS,
       author = {{Cheng}, Quan and {Yu}, Yun-Wei and {Zheng}, Xiao-Ping},
        title = "{Stochastic gravitational wave background from magnetic deformation of newly born magnetars}",
      journal = {\mnras},
         year = 2015,
        month = dec,
       volume = {454},
       number = {3},
        pages = {2299-2304},
          doi = {10.1093/mnras/stv2127},
archivePrefix = {arXiv},
       eprint = {1509.07651},
 primaryClass = {astro-ph.SR},
       adsurl = {https://ui.adsabs.harvard.edu/abs/2015MNRAS.454.2299C}
}

@ARTICLE{Yan2024EPJC,
       author = {{Yan}, Yu-Long and {Cheng}, Quan and {Zheng}, Xiao-Ping and {Ouyang}, Xia-Xia},
        title = "{On the initial spin periods of magnetars born in weak supernova explosions and their gravitational wave radiation}",
      journal = {European Physical Journal C},
         year = 2024,
        month = oct,
       volume = {84},
       number = {10},
          eid = {1043},
        pages = {1043},
          doi = {10.1140/epjc/s10052-024-13406-0},
archivePrefix = {arXiv},
       eprint = {2410.23576},
 primaryClass = {astro-ph.HE},
       adsurl = {https://ui.adsabs.harvard.edu/abs/2024EPJC...84.1043Y}
}

@ARTICLE{Sathyaprakash2009LRR,
       author = {{Sathyaprakash}, B.~S. and {Schutz}, Bernard F.},
        title = "{Physics, Astrophysics and Cosmology with Gravitational Waves}",
      journal = {Living Reviews in Relativity},
         year = 2009,
        month = dec,
       volume = {12},
       number = {1},
          eid = {2},
        pages = {2},
          doi = {10.12942/lrr-2009-2},
archivePrefix = {arXiv},
       eprint = {0903.0338},
 primaryClass = {gr-qc},
       adsurl = {https://ui.adsabs.harvard.edu/abs/2009LRR....12....2S}
}

@ARTICLE{Yagi2011PhRvD,
       author = {{Yagi}, Kent and {Seto}, Naoki},
        title = "{Detector configuration of DECIGO/BBO and identification of cosmological neutron-star binaries}",
      journal = {\prd},
         year = 2011,
        month = feb,
       volume = {83},
       number = {4},
          eid = {044011},
        pages = {044011},
          doi = {10.1103/PhysRevD.83.044011},
archivePrefix = {arXiv},
       eprint = {1101.3940},
 primaryClass = {astro-ph.CO},
       adsurl = {https://ui.adsabs.harvard.edu/abs/2011PhRvD..83d4011Y}
}

@ARTICLE{Thompson1993ApJ,
       author = {{Thompson}, Christopher and {Duncan}, Robert C.},
        title = "{Neutron Star Dynamos and the Origins of Pulsar Magnetism}",
      journal = {\apj},
         year = 1993,
        month = may,
       volume = {408},
        pages = {194},
          doi = {10.1086/172580},
       adsurl = {https://ui.adsabs.harvard.edu/abs/1993ApJ...408..194T}
}

@ARTICLE{Duncan1992ApJ,
       author = {{Duncan}, Robert C. and {Thompson}, Christopher},
        title = "{Formation of Very Strongly Magnetized Neutron Stars: Implications for Gamma-Ray Bursts}",
      journal = {\apjl},
         year = 1992,
        month = jun,
       volume = {392},
        pages = {L9},
          doi = {10.1086/186413},
       adsurl = {https://ui.adsabs.harvard.edu/abs/1992ApJ...392L...9D}
}

@ARTICLE{Kasen2010ApJ,
       author = {{Kasen}, Daniel and {Bildsten}, Lars},
        title = "{Supernova Light Curves Powered by Young Magnetars}",
      journal = {\apj},
         year = 2010,
        month = jul,
       volume = {717},
       number = {1},
        pages = {245-249},
          doi = {10.1088/0004-637X/717/1/245},
archivePrefix = {arXiv},
       eprint = {0911.0680},
 primaryClass = {astro-ph.HE},
       adsurl = {https://ui.adsabs.harvard.edu/abs/2010ApJ...717..245K}
}

@ARTICLE{Bernardini2015JHEAp,
       author = {{Bernardini}, Maria Grazia},
        title = "{Gamma-ray bursts and magnetars: Observational signatures and predictions}",
      journal = {Journal of High Energy Astrophysics},
         year = 2015,
        month = sep,
       volume = {7},
        pages = {64-72},
          doi = {10.1016/j.jheap.2015.05.003},
       adsurl = {https://ui.adsabs.harvard.edu/abs/2015JHEAp...7...64B}
}

@ARTICLE{Vink2006MNRAS,
       author = {{Vink}, Jacco and {Kuiper}, Lucien},
        title = "{Supernova remnant energetics and magnetars: no evidence in favour of millisecond proto-neutron stars}",
      journal = {\mnras},
         year = 2006,
        month = jul,
       volume = {370},
       number = {1},
        pages = {L14-L18},
          doi = {10.1111/j.1745-3933.2006.00178.x},
archivePrefix = {arXiv},
       eprint = {astro-ph/0604187},
 primaryClass = {astro-ph},
       adsurl = {https://ui.adsabs.harvard.edu/abs/2006MNRAS.370L..14V}
}

@ARTICLE{Jawor2022MNRAS,
       author = {{Jawor}, Jedrzej A. and {Tauris}, Thomas M.},
        title = "{Modelling spin evolution of magnetars}",
      journal = {\mnras},
         year = 2022,
        month = jan,
       volume = {509},
       number = {1},
        pages = {634-657},
          doi = {10.1093/mnras/stab2677},
archivePrefix = {arXiv},
       eprint = {2109.07484},
 primaryClass = {astro-ph.HE},
       adsurl = {https://ui.adsabs.harvard.edu/abs/2022MNRAS.509..634J}
}

@ARTICLE{Huang2024PhRvD,
       author = {{Huang}, Jun-Xiang and {L{\"u}}, Hou-Jun and {Rice}, Jared and {Liang}, En-Wei},
        title = "{Spin, inclination, and magnetic field evolution of magnetar population in vacuum and plasma-filled magnetospheres}",
      journal = {\prd},
         year = 2024,
        month = jun,
       volume = {109},
       number = {12},
          eid = {123026},
        pages = {123026},
          doi = {10.1103/PhysRevD.109.123026},
archivePrefix = {arXiv},
       eprint = {2405.15484},
 primaryClass = {astro-ph.HE},
       adsurl = {https://ui.adsabs.harvard.edu/abs/2024PhRvD.109l3026H}
}

@ARTICLE{Aguilera2008AA,
       author = {{Aguilera}, D.~N. and {Pons}, J.~A. and {Miralles}, J.~A.},
        title = "{2D Cooling of magnetized neutron stars}",
      journal = {\aap},
         year = 2008,
        month = jul,
       volume = {486},
       number = {1},
        pages = {255-271},
          doi = {10.1051/0004-6361:20078786},
archivePrefix = {arXiv},
       eprint = {0710.0854},
 primaryClass = {astro-ph},
       adsurl = {https://ui.adsabs.harvard.edu/abs/2008A&A...486..255A}
}

@ARTICLE{Popov2012MNRAS,
       author = {{Popov}, S.~B. and {Turolla}, R.},
        title = "{Probing the neutron star spin evolution in the young Small Magellanic Cloud Be/X-ray binary SXP 1062}",
      journal = {\mnras},
         year = 2012,
        month = mar,
       volume = {421},
       number = {1},
        pages = {L127-L131},
          doi = {10.1111/j.1745-3933.2012.01220.x},
archivePrefix = {arXiv},
       eprint = {1112.2507},
 primaryClass = {astro-ph.HE},
       adsurl = {https://ui.adsabs.harvard.edu/abs/2012MNRAS.421L.127P}
}

@ARTICLE{Zhou2024MNRAS,
       author = {{Zhou}, Xia and {Huang}, Hai-Tao and {Cheng}, Quan and {Zheng}, Xiao-Ping},
        title = "{Formation of long-period radio pulsars}",
      journal = {\mnras},
         year = 2024,
        month = may,
       volume = {530},
       number = {2},
        pages = {1636-1643},
          doi = {10.1093/mnras/stae954},
       adsurl = {https://ui.adsabs.harvard.edu/abs/2024MNRAS.530.1636Z}
}

@book{Maggioregw,
    author = {Maggiore, Michele},
    title = {Gravitational Waves: Volume 1: Theory and Experiments},
    publisher = {Oxford University Press},
    year = {2007},
    month = {10},
    isbn = {9780198570745},
    doi = {10.1093/acprof:oso/9780198570745.001.0001},
    url = {https://doi.org/10.1093/acprof:oso/9780198570745.001.0001},
}

@ARTICLE{Esposito2013MNRAS,
       author = {{Esposito}, P. and {Tiengo}, A. and {Rea}, N. and {Turolla}, R. and {Fenzi}, A. and {Giuliani}, A. and {Israel}, G.~L. and {Zane}, S. and {Mereghetti}, S. and {Possenti}, A. and {Burgay}, M. and {Stella}, L. and {G{\"o}tz}, D. and {Perna}, R. and {Mignani}, R.~P. and {Romano}, P.},
        title = "{X-ray and radio observations of the magnetar Swift J1834.9-0846 and its dust-scattering halo}",
      journal = {\mnras},
         year = 2013,
        month = mar,
       volume = {429},
       number = {4},
        pages = {3123-3132},
          doi = {10.1093/mnras/sts569},
archivePrefix = {arXiv},
       eprint = {1212.1079},
 primaryClass = {astro-ph.HE},
       adsurl = {https://ui.adsabs.harvard.edu/abs/2013MNRAS.429.3123E}
}

@ARTICLE{Geppert2014MNRAS,
       author = {{Geppert}, U. and {Vigan{\`o}}, D.},
        title = "{Creation of magnetic spots at the neutron star surface}",
      journal = {\mnras},
         year = 2014,
        month = nov,
       volume = {444},
       number = {4},
        pages = {3198-3208},
          doi = {10.1093/mnras/stu1675},
archivePrefix = {arXiv},
       eprint = {1408.3833},
 primaryClass = {astro-ph.SR},
       adsurl = {https://ui.adsabs.harvard.edu/abs/2014MNRAS.444.3198G}
}

@ARTICLE{Passamonti2017MNRAS,
       author = {{Passamonti}, Andrea and {Akg{\"u}n}, Taner and {Pons}, Jos{\'e} A. and {Miralles}, Juan A.},
        title = "{The relevance of ambipolar diffusion for neutron star evolution}",
      journal = {\mnras},
         year = 2017,
        month = mar,
       volume = {465},
       number = {3},
        pages = {3416-3428},
          doi = {10.1093/mnras/stw2936},
archivePrefix = {arXiv},
       eprint = {1608.00001},
 primaryClass = {astro-ph.HE},
       adsurl = {https://ui.adsabs.harvard.edu/abs/2017MNRAS.465.3416P}
}

@ARTICLE{Glampedakis2011MNRAS,
       author = {{Glampedakis}, K. and {Jones}, D.~I. and {Samuelsson}, L.},
        title = "{Ambipolar diffusion in superfluid neutron stars}",
      journal = {\mnras},
         year = 2011,
        month = may,
       volume = {413},
       number = {3},
        pages = {2021-2030},
          doi = {10.1111/j.1365-2966.2011.18278.x},
archivePrefix = {arXiv},
       eprint = {1010.1153},
 primaryClass = {astro-ph.SR},
       adsurl = {https://ui.adsabs.harvard.edu/abs/2011MNRAS.413.2021G}
}

@ARTICLE{Kantor2018MNRAS,
       author = {{Kantor}, E.~M. and {Gusakov}, M.~E.},
        title = "{A note on the ambipolar diffusion in superfluid neutron stars}",
      journal = {\mnras},
         year = 2018,
        month = jan,
       volume = {473},
       number = {3},
        pages = {4272-4277},
          doi = {10.1093/mnras/stx2682},
archivePrefix = {arXiv},
       eprint = {1703.09216},
 primaryClass = {astro-ph.HE},
       adsurl = {https://ui.adsabs.harvard.edu/abs/2018MNRAS.473.4272K}
}

@ARTICLE{Jones1988MNRAS,
       author = {{Jones}, P.~B.},
        title = "{Neutron star magnetic field decay - Hall drift and Ohmic diffusion}",
      journal = {\mnras},
         year = 1988,
        month = aug,
       volume = {233},
        pages = {875-885},
          doi = {10.1093/mnras/233.4.875},
       adsurl = {https://ui.adsabs.harvard.edu/abs/1988MNRAS.233..875J}
}

@ARTICLE{Cerri-Serim2019MNRAS,
       author = {{{\'C}erri-Serim}, D. and {Serim}, M.~M. and {{\c{S}}ahiner}, {\c{S}}. and {Inam}, S. {\'c}. and {Baykal}, A.},
        title = "{Pulse frequency fluctuations of magnetars}",
      journal = {\mnras},
         year = 2019,
        month = may,
       volume = {485},
       number = {1},
        pages = {2-12},
          doi = {10.1093/mnras/sty3213},
archivePrefix = {arXiv},
       eprint = {1806.00401},
 primaryClass = {astro-ph.HE},
       adsurl = {https://ui.adsabs.harvard.edu/abs/2019MNRAS.485....2C}
}

@ARTICLE{Tsang2013ApJ,
       author = {{Tsang}, David and {Gourgouliatos}, Konstantinos N.},
        title = "{Timing Noise in Pulsars and Magnetars and the Magnetospheric Moment of Inertia}",
      journal = {\apjl},
         year = 2013,
        month = aug,
       volume = {773},
       number = {1},
          eid = {L17},
        pages = {L17},
          doi = {10.1088/2041-8205/773/1/L17},
archivePrefix = {arXiv},
       eprint = {1302.4448},
 primaryClass = {astro-ph.HE},
       adsurl = {https://ui.adsabs.harvard.edu/abs/2013ApJ...773L..17T}
}

@ARTICLE{Ashton2015PhRvD,
       author = {{Ashton}, G. and {Jones}, D.~I. and {Prix}, R.},
        title = "{Effect of timing noise on targeted and narrow-band coherent searches for continuous gravitational waves from pulsars}",
      journal = {\prd},
         year = 2015,
        month = mar,
       volume = {91},
       number = {6},
          eid = {062009},
        pages = {062009},
          doi = {10.1103/PhysRevD.91.062009},
archivePrefix = {arXiv},
       eprint = {1410.8044},
 primaryClass = {gr-qc},
       adsurl = {https://ui.adsabs.harvard.edu/abs/2015PhRvD..91f2009A}
}

@ARTICLE{Ashton2018PhRvD,
       author = {{Ashton}, G. and {Prix}, R. and {Jones}, D.~I.},
        title = "{A semicoherent glitch-robust continuous-gravitational-wave search method}",
      journal = {\prd},
         year = 2018,
        month = sep,
       volume = {98},
       number = {6},
          eid = {063011},
        pages = {063011},
          doi = {10.1103/PhysRevD.98.063011},
archivePrefix = {arXiv},
       eprint = {1805.03314},
 primaryClass = {gr-qc},
       adsurl = {https://ui.adsabs.harvard.edu/abs/2018PhRvD..98f3011A}
}

@ARTICLE{Keitel2019PhRvD,
       author = {{Keitel}, David and {Woan}, Graham and {Pitkin}, Matthew and {Schumacher}, Courtney and {Pearlstone}, Brynley and {Riles}, Keith and {Lyne}, Andrew G. and {Palfreyman}, Jim and {Stappers}, Benjamin and {Weltevrede}, Patrick},
        title = "{First search for long-duration transient gravitational waves after glitches in the Vela and Crab pulsars}",
      journal = {\prd},
         year = 2019,
        month = sep,
       volume = {100},
       number = {6},
          eid = {064058},
        pages = {064058},
          doi = {10.1103/PhysRevD.100.064058},
archivePrefix = {arXiv},
       eprint = {1907.04717},
 primaryClass = {gr-qc},
       adsurl = {https://ui.adsabs.harvard.edu/abs/2019PhRvD.100f4058K}
}

@ARTICLE{Suvorova2016PhRvD,
       author = {{Suvorova}, S. and {Sun}, L. and {Melatos}, A. and {Moran}, W. and {Evans}, R.~J.},
        title = "{Hidden Markov model tracking of continuous gravitational waves from a neutron star with wandering spin}",
      journal = {\prd},
         year = 2016,
        month = jun,
       volume = {93},
       number = {12},
          eid = {123009},
        pages = {123009},
          doi = {10.1103/PhysRevD.93.123009},
archivePrefix = {arXiv},
       eprint = {1606.02412},
 primaryClass = {astro-ph.IM},
       adsurl = {https://ui.adsabs.harvard.edu/abs/2016PhRvD..93l3009S}
}

@ARTICLE{Ge2025SCPMA,
       author = {{Ge}, Mingyu and {Ji}, Long and {Taverna}, Roberto and {Tsygankov}, Sergey and {Xu}, Yanjun and {Santangelo}, Andrea and {Zane}, Silvia and {Zhang}, Shuang-Nan and {Feng}, Hua and {Chen}, Wei and {Cheng}, Quan and {Hou}, Xian and {Imbrogno}, Matteo and {Israel}, Gian Luca and {Kelly}, Ruth and {Kong}, Ling-Da and {Liu}, Kuan and {Mushtukov}, Alexander and {Poutanen}, Juri and {Suleimanov}, Valery and {Tao}, Lian and {Tong}, Hao and {Turolla}, Roberto and {Wang}, Weihua and {Ye}, Wentao and {Zhao}, Qing-Chang and {Brice}, Nabil and {Geng}, Jinjun and {Lin}, Lin and {Wang}, Wei-Yang and {Xie}, Fei and {Xiong}, Shao-Lin and {Zhang}, Shu and {Fu}, Yucong and {Lai}, Dong and {Li}, Jian and {Li}, Pan-Ping and {Li}, Xiaobo and {Li}, Xinyu and {Liu}, Honghui and {Liu}, Jiren and {Peng}, Jingqiang and {Shui}, Qingcang and {Tuo}, Youli and {Wang}, Hongguang and {Wang}, Wei and {Weng}, Shanshan and {You}, Yuan and {Zheng}, Xiaoping and {Zhou}, Xia},
        title = "{Physics of strong magnetism with eXTP}",
      journal = {Science China Physics, Mechanics, and Astronomy},
         year = 2025,
        month = sep,
       volume = {68},
       number = {11},
          eid = {119505},
        pages = {119505},
          doi = {10.1007/s11433-025-2796-y},
archivePrefix = {arXiv},
       eprint = {2506.08369},
 primaryClass = {astro-ph.HE},
       adsurl = {https://ui.adsabs.harvard.edu/abs/2025SCPMA..6819505G}
}
\bibliographystyle{aasjournal}

\label{lastpage}
\end{CJK*}
\end{document}